\documentclass[12pt,a4paper]{article}
\usepackage{amsfonts}
\usepackage{mathrsfs}  
\usepackage{epsfig}
\usepackage{axodraw}
\usepackage{graphicx}

\def\simlt{\stackrel{<}{{}_\sim}}

\def\lrvec#1{\vbox{\ialign{##\crcr
${\hspace{1pt}\scriptscriptstyle\leftrightarrow\hspace{-1pt}}
$\crcr\noalign{\nointerlineskip}
$\hfil\displaystyle{#1}\hfil$\crcr}}}

\begin{document}

\title{Ground state energy of the polarized diluted gas of interacting spin $1/2$ fermions}
\author{\em Piotr Chankowski and Jacek Wojtkiewicz\footnote{Emails:
    chank@fuw.edu.pl, wjacek@fuw.edu.pl} \\
Faculty of Physics, University of Warsaw,\\
Pasteura 5, 02-093 Warszawa, Poland
}
\maketitle
\abstract{The effective field theory approach simplifies the perturbative
computation of the ground state energy of the diluted gas of repulsive
fermions allowing in the case of the unpolarized system to easily rederive
the classic results up to the $(k_{\rm F}a_0)^2$ order (where $k_{\rm F}$ is
the system's Fermi momentum and $a_0$ the $s$-wave scattering length) and
(with more labour) to extend it up to the order $(k_{\rm F}a_0)^4$. The
analogous expansion of the ground state energy of the polarized gas of
spin $1/2$ fermions is known only up to the $k_{\rm F}a_0$ order (where
$k_{\rm F}$ stands for $k_{{\rm F}\uparrow}$ or $k_{{\rm F}\downarrow}$); the
order $(k_{\rm F}a_0)^2$ contribution has been computed (analytically)
only using a specific (hard-core type) interaction potential. Here
we show that the effective field theory method also allows to easily
obtain the order $(k_{\rm F}a_0)^2$ correction to the ground state of
the polarized gas in a way applicable to all repulsive interactions.
\vskip0.1cm

}

\newpage

\section{Introduction}
\label{sec:introd}

Effective field theories are used in high energy physics since already
half a century. They allow for example to quantitatively capture,
by exploiting essentially only the information about exact and broken
symmetries of the underlying theory, characteristic features of the
low energy dynamics of light mesons in the regime in which quantum
chromodynamics is genuinely nonperturbative and, on the other hand,
routinely serve to parametrize potential effects of yet unknown new physics
in processes involving well known particles. These applications rely on
the separation of energy scales involved which makes reliable the
expansion of the computed quantities in powers of their ratio and on
the possibility of fixing values of the parameters, which cannot be
obtained by matching onto the underlying more fundamental
theory, by directly extracting them from low energy data.

Relatively more recent are applications of the effective field theory
methods to nonrelativistic many body problems
\cite{KolckiSka}--\cite{Mloty3}. A particularly instructive
is the application of this technique \cite{HamFur00} to the classic problem
of computing the energy $E_\Omega$ of the ground state of the system of $N$
fermions (enclosed in the volume $V$) interacting through a two-body spin
independent potential which may not be specified explicitly but is, instead,
characterized by the (in principle infinite) set of the scattering lengths
$a_\ell$ and the effective radii $r_\ell$ ($\ell=0,1,\dots$) parametrizing the
expansion of the resulting partial amplitudes of elastic scattering of
two fermions in powers of their relative momentum. Stated in this form the
problem is ideally suited for handling it in the framework of an effective
theory, because the information on the fundamental dynamics (the two-body
potential) is traded from the beginning for the (infinite) set of low
energy data. The crucial observation making possible the application of the
effective theory technique to it is that if the underlying interaction
potential is natural in the sense that the magnitudes of the scattering
lengths $a_\ell$ and the effective radii $r_\ell$ it gives rise to are set by
some common scale $1/\Lambda$ (this excludes from the considerations attractive
potentials which can lead to bound states and formation of resonances),
this scale is, if the gas of fermions is sufficiently diluted, well separated
from its characteristic momentum scale set by the Fermi momentum (wave vector)
$k_{\rm F}$ of the gas. The expansion of the ground state energy $E_\Omega/N$
or $E_\Omega/V$ in powers of $k_{\rm F}a_\ell\propto k_{\rm F}/\Lambda$ and
$k_{\rm F}r_\ell\propto k_{\rm F}/\Lambda$
naturally provided by the effective theory methods is then reliable.
In the textbook treatment of this problem \cite{FetWal} which summarizes
results obtained in the past within the conventional approaches
(mean field approach \cite{Lenz}, ordinary perturbative expansion
\cite{HuangYang57}, Goldstone diagrams, \cite{dDM} or Green's functions
methods \cite{Baker}-\cite{Amusia68}) computing already the the term of
order ${\cal O}(k^2_{\rm F}a_0^2)$ in the expansion of the energy $E_\Omega$ of
the gas of unpolarized fermions requires a sophisticated procedure consisting
of summation of an infinite subset of Goldstone diagrams and re-expanding
the resulting expressions. In contrast, in the effective field theory
determination of the term of order ${\cal O}(k^2_{\rm F}a_0^2)$ reduces
to a computation of a single (divergent) diagram and removing its divergence
with the help the standard renormalization procedure.
Owing to its simplicity this method allowed to obtain \cite{WeDrSch} recently
the complete fourth (${\cal O}(k^4_{\rm F}/\Lambda^4)$ and
${\cal O}((k^4_{\rm F}/\Lambda^4)\ln(k_{\rm F}/\Lambda)$) order terms which
complement the
third, ${\cal O}(k^3_{\rm F}/\Lambda^3)$, order result obtained earlier
\cite{Kaiser} by more conventional (semi-analytic) methods.

The interest in properties of a diluted gas of fermions stemmed originally
from the study of nuclear matter, although this model obviously cannot
capture all realistic features of systems of nucleons interacting through
(mostly) attractive potentials. More recently models of diluted gases
(of fermions and bosons) find their more natural application as continuum
models of interacting cold atomic gases bound in optical or harmonic
traps, complementing more traditional ways of investigating properties
of such systems based on lattice models known generally as Hubbard models
(paradigmatic for condensed matter physics) which, despite of more than
60 years of development, still leave many problems without clear answers.
One of them is the mechanism of emergence of 
ferromagnetism in
systems of mutually repelling atoms which has been observed experimentally
\cite{Jo}. Theoretical investigations of many questions of interest
related to this result, like the problem of itinerant ferromagnetism
on lattice models of mutually repelling spin 1/2 fermions as well as
the possibility of spontaneous separation of magnetic and nonmagnetic
phases begun already in the sixtieth of the XX century, but a successful
explanation of these phenomena with the help of the so-called dynamical
mean field theory (DMFT) approach \cite{MetallicFM} has been achieved
only some twenty years ago.

Clearly, any study of the emergence of magnetism in systems of $N$ mutually
repelling spin 1/2 fermions within the continuum model must start with the
computation of the ground state energy of such a system for
$N_\uparrow\neq N_\downarrow$, where $N_\uparrow$ and $N_\downarrow$
($N_\uparrow+ N_\downarrow=N$) are the (conserved by the assumed interaction)
numbers of spin up and spin down fermions. Using the ordinary
Rayleigh-Schr\"odinger perturbative expansion it is easy to obtain the
relevant expression in the first order in the $s$-wave scattering length
$a_0$ (this result can be given also a mathematically more rigorous
foundation \cite{LSS}, \cite{Porta}). The second order term has been
computed analytically using the traditional approach only within the hard
spheres model interaction \cite{KANNO}. This approach does not allow, however,
to easily recognize the universality of this result (in the class of natural
spin-independent repulsive potentials). Apart from these result, there are
also Monte Carlo simulations \cite{QMC10}, \cite{QMC14} which -- while
providing quite reliable numerical estimates of the exact (nonperturbative)
ground state energy --  must necessarily employ concrete model potentials and
therefore suffer from the lack of universality.

In this paper we compute the second order correction to the ground state
energy of the polarized gas of spin $1/2$ fermions with the help of the
effective field theory method. It makes it clear from the outset that
this correction can only depend on the $s$-wave scattering length $a_0$ and,
therefore, that the result of \cite{KANNO} is universal. Nevertheless, it
is interestig to recover it using this new method as this may pave the way
to extend the computation to yet higher orders. From the conceptual point
of view, this task reduces to only a minor modification of the computation
performed in \cite{HamFur00} for $N_\uparrow=N_\downarrow$, but it is
 more involved from technical point of view.
While the order $a_0^2$ correction to the ground state energy is in
\cite{HamFur00} given in a completely analytic form, and, as the result of
\cite{KANNO} shows, the same is possible also in the case of
$N_\uparrow\neq N_\downarrow$, we do not attempt to perform the resulting
integrals analytically and content ourselves with providing the
formulae which involve integrals which can be easily evaluated numerically
with the help of a three-line Mathematica code.

Since our computation parallels that of \cite{HamFur00}, we take the
opportunity to give here more details (and to make some general
comments on the renormalization procedure) which, although probably
obvious to experts, may be of some pedagogical value. Therefore in Section
\ref{sec:zerothandfirst} we state first the problem in the formalism of
second quantization and compute by the standard perturbative
method the first (order $a_0$) correction to the ground state energy
of the system of interacting  fermions. The purpose of this section is
also to fix the notation and possible ways of parametrizing the result.
In Section \ref{sec:efftheory} we recall the effective field theory approach
and give some details of the procedure allowing to relate the couplings
of the effective Lagrangian (Hamiltonian) to the ``low energy data'',
that is, to the scattering lengths $a_\ell$ and effective radii $r_\ell$.
Instead of using as in \cite{HamFur00} the dimensional reduction as the
regularization method, we cut off divergent integrals over the wave vectors
at the scale $\Lambda$. While being technically more troublesome
(but only in higher orders) this regularization prescription seems more in
line with the main idea of the effective theory method and moreover it
allows to partly control the correctness of the calculation, which is not
possible with the dimensional regularization which automatically sets
to zero all power-like divergences.
In Section \ref{sec:Ecorrections} we compute the second-order correction to
the ground state energy using the effective theory method, demonstrate
explicitly cancellation of the dependence on the cutoff $\Lambda$
and give the result in the form dependent on a single function of
the ratio of Fermi momenta of spin up and spin down fermions
which is evaluated numerically. We also compare our perturbative result with
the existing nonperturbative estimates 
based on Monte Carlo simulations. 
We summarize the results in Section \ref{sec:summary}
and speculate about perspectives of generalizing them.

\section{Zeroth and first order results}
\label{sec:zerothandfirst}

\noindent We would like to compute energy $E_\Omega$ of the ground state
of $N$ identical (nonrelativistic) fermions of mass $m_f$ and spin $s=1/2$
($g=2s+1=2$ spin states per fermion) interacting through a spin independent
two-body potential which, instead of being specified explicitly, is
characterized in terms of the scattering lengths and effective ranges
of the elastic scattering amplitude it gives rise to. We will not
assume equal densities $\rho_\uparrow=N_\uparrow/V$ of fermions having
spin up and  $\rho_\downarrow=N_\downarrow/V$  ($N_\uparrow+N_\downarrow=N$)
of fermions having spin down.

We first assume that the interaction potential is given explicitly and 
the Hamiltonian of the system (enclosed in the box of volume $V$ and with
periodic boundary conditions imposed) in the standard formalism of second
quantization (see e.g. \cite{FetWal,Feynman}) has the usual form
$H=H_0+V_{\rm int}$ with
\begin{eqnarray}
  H_0=\sum_{\mathbf{p},\sigma=\uparrow,\downarrow}{\hbar^2\mathbf{p}^2\over2m_f}~\!
  a^\dagger_{\mathbf{p},\sigma} a_{\mathbf{p},\sigma}~\!,
  \phantom{aaaaaaaaaaaaaaaaaaaaaaaaaaaaa}~\!\label{eqn:ModelHamiltonian}\\
  V_{\rm int}={1\over2V}\sum_{\mathbf{q}}\tilde V_{\rm pot}(\mathbf{q})\!
  \sum_{\mathbf{p}_1,\mathbf{p}_2}
  \sum_{\sigma_1,\sigma_2=\uparrow,\downarrow}a^\dagger_{\mathbf{p}_1+\mathbf{q},\sigma_1}
  a^\dagger_{\mathbf{p}_2-\mathbf{q},\sigma_2}a_{\mathbf{p}_2,\sigma_2}a_{\mathbf{p}_1,\sigma_1}~\!,
  \label{eqn:ModelInteraction}
\end{eqnarray}
where $\tilde V_{\rm pot}(\mathbf{q})$ is the Fourier transform of the
assumed spin-independent two-body interaction potential
$V_{\rm pot}(\mathbf{r}_2-\mathbf{r}_1)$.
The Hamiltonian $H$ is ordered normally
with respect to the lowest vector $|{\rm void}\rangle$ of the Fock space
(in which act the creation $a^\dagger_{\mathbf{p},\sigma}$ and annihilation
$a_{\mathbf{p},\sigma}$ operators); this ensures equivalence of the second
quantization formalism with the one based on the $N$-body Schr\"odinger
wave equation. The fixed numbers $N_\uparrow$ and $N_\downarrow$ determine
the Fermi momenta $p_{{\rm F}\uparrow}$ and $p_{{\rm F}\downarrow}$ through the
relations\footnote{The standard replacement
  $\sum_{\mathbf{k}}\rightarrow V\int\!d^3\mathbf{k}/(2\pi)^3$ is assumed.}
\begin{eqnarray}
  N_{\uparrow/\downarrow}=
  V\!\int\!{d^3\mathbf{k}\over(2\pi)^3}~\!
  \theta\!\left(p_{{\rm F}\uparrow/\downarrow}-|\mathbf{k}|\right)={V\over6\pi^2}~\!
  p^3_{{\rm F}\uparrow/\downarrow}~\!.
\end{eqnarray}
We seek the ground state energy $E_\Omega$ of $H$ in the form of the series
$E_\Omega=E_{\Omega_0}+E_\Omega^{(1)}+\dots$ The first term $E_\Omega^{(0)}=E_{\Omega_0}$
is the energy of the ground state $|\Omega_0\rangle$ of the system of $N$
noninteracting fermions
\begin{eqnarray}
  E_{\Omega_0}
  =V\!\sum_{s=\uparrow,\downarrow}\int\!{d^3\mathbf{k}\over(2\pi)^3}~\!
  {\hbar^2\mathbf{p}^2\over2m_f}~\!
  \theta\!\left(p_{{\rm F}s}-|\mathbf{k}|\right)
  ={V\over6\pi^2}~\!{3\over5}~\!{\hbar^2\over2m_f}\left(p^5_{{\rm F}\uparrow}
  +p^5_{{\rm F}\downarrow}\right)\phantom{a}\nonumber\\
  ={3\over5}\left(N_\uparrow~\!
  {\hbar^2p^2_{{\rm F}\uparrow}\over2m_f}+N_\downarrow~\!
  {\hbar^2p^2_{{\rm F}\downarrow}\over2m_f}\right)={3\over5}~\!{\hbar^2\over2m_f}~\!
  (6\pi^2)^{2/3}\left(\rho_\uparrow^{5/3}+\rho_\downarrow^{5/3}\right),\nonumber
\end{eqnarray}
where $\rho_{\uparrow/\downarrow}=N_{\uparrow/\downarrow}/V$.

The first order correction $E^{(1)}_\Omega$ can be computed in
different ways. Application of the standard Rayleigh-Schr\"odinger
perturbative expansion gives it in the form of the matrix element
\begin{eqnarray}
  E^{(1)}_\Omega={1\over2V}\sum_{\mathbf{q}}\tilde V_{\rm pot}(\mathbf{q})
  \sum_{\mathbf{p}_1,\mathbf{p}_2}
  \sum_{\sigma_1,\sigma_2=\uparrow,\downarrow}\langle\Omega_0|a^\dagger_{\mathbf{p}_1+\mathbf{q},\sigma_1}
  a^\dagger_{\mathbf{p}_2-\mathbf{q},\sigma_2}a_{\mathbf{p}_2,\sigma_2}a_{\mathbf{p}_1,\sigma_1}
  |\Omega_0\rangle~\!,\nonumber
\end{eqnarray}
which can be straightforwardly evaluated by applying the Wick theorem
(\cite{FetWal,Molinari}) to the string of two creation and two annihilation
operators
\begin{eqnarray}
  a^\dagger_{\mathbf{p}_1+\mathbf{q},\sigma_1}
  a^\dagger_{\mathbf{p}_2-\mathbf{q},\sigma_2}a_{\mathbf{p}_2,\sigma_2}a_{\mathbf{p}_1,\sigma_1}
  =[a^\dagger_{\mathbf{p}_1+\mathbf{q},\sigma_1}a_{\mathbf{p}_1,\sigma_1}]
  [a^\dagger_{\mathbf{p}_2-\mathbf{q},\sigma_2}a_{\mathbf{p}_2,\sigma_2}]\phantom{aaaaa}~\nonumber\\
  -~\![a^\dagger_{\mathbf{p}_1+\mathbf{q},\sigma_1}a_{\mathbf{p}_2,\sigma_2}]
  [a^\dagger_{\mathbf{p}_2-\mathbf{q},\sigma_2}a_{\mathbf{p}_1,\sigma_1}]\phantom{aaaaa}~\nonumber\\
  +~\![a^\dagger_{\mathbf{p}_1+\mathbf{q},\sigma_1}
  a^\dagger_{\mathbf{p}_2-\mathbf{q},\sigma_2}][a_{\mathbf{p}_2,\sigma_2}a_{\mathbf{p}_1,\sigma_1}]
  +\dots,\nonumber
\end{eqnarray}
where the ellipsis stands for operator terms ordered normally with respect to
the state $|\Omega_0\rangle$ (characterized by $p_{{\rm F}\uparrow}$ and
$p_{{\rm F}\downarrow}$) of $N$ noninteracting fermions, and the pairings
$[a^\dagger_{\mathbf{p}_1+\mathbf{q},\sigma_1}a_{\mathbf{p}_1,\sigma_1}]$  etc.
are given by (pairings of $aa$ and  $a^\dagger a^\dagger$ vanish)
\begin{eqnarray}
  [a^\dagger_{\mathbf{p},\sigma}a_{\mathbf{p}^\prime,\sigma^\prime}]=
  \langle\Omega_0|a^\dagger_{\mathbf{p},\sigma_1}a_{\mathbf{p}^\prime,\sigma^\prime}|
\Omega_0\rangle=
  \delta_{\sigma^\prime\sigma}\delta_{\mathbf{p},\mathbf{p}^\prime}
  \theta(p_{{\rm F}\sigma}-|\mathbf{p}|)~\!\theta(p_{{\rm F}\sigma}-|\mathbf{p}^\prime|)~\!.
  \nonumber
\end{eqnarray}
The contributions to $E_\Omega^{(1)}$ of the terms in which all operators have spin
up or all operators have spin down are proportional to
\begin{eqnarray}
  \sum_{\mathbf{p}_1,\mathbf{p}_2}\left(\tilde V_{\rm pot}(\mathbf{0})
  -\tilde V_{\rm pot}(\mathbf{p}_2-\mathbf{p}_1)\right)
  \theta(p_{\rm F}-|\mathbf{p}_1|)
  \theta(p_{\rm F}-|\mathbf{p}_2|)~\!,\nonumber
\end{eqnarray}
and vanish in the approximation $\mathbf{p}_2-\mathbf{p}_1\sim\mathbf{0}$
which is valid if the gas is diluted so that both $p_{{\rm F}\uparrow}$
and $p_{{\rm F}\downarrow}$ are small. The two remaining terms give equal
contributions and one obtains
\begin{eqnarray}
  E_\Omega^{(1)}={1\over V}~\!\tilde V_{\rm pot}(\mathbf{0})
  \sum_{\mathbf{p}_1}\theta(p_{{\rm F}\uparrow}-|\mathbf{p}_1|)
  \sum_{\mathbf{p}_2}\theta(p_{{\rm F}\downarrow}-|\mathbf{p}_1|)
  =V~\!\tilde V_{\rm pot}(\mathbf{0})~\!{p^3_{{\rm F}\uparrow}\over6\pi^2}
  ~\!{p^3_{{\rm F}\downarrow}\over6\pi^2}~\!.
\end{eqnarray}
Even if fermions interact via  a hardcore-type potential
$V_{\rm pot}(|\mathbf{x}_1-\mathbf{x}_2|)$ the Fourier transform of
which may be ill defined, the correct result, as argued e.g.
in \cite{FetWal}, is obtained by trading $\tilde V_{\rm pot}(\mathbf{0})$
for the $s$-wave scattering length $a_0$ (which is
defined as the $k\rightarrow0$ limit of the always
well defined scattering amplitude) according to the formula
\begin{eqnarray}
  \tilde V_{\rm pot}(\mathbf{0})=\int\!d^3\mathbf{x}~\!V_{\rm pot}(\mathbf{x})
   =\lim_{k\rightarrow0}\left(-{2\pi\hbar^2\over m_{\rm red}}~\!
   f_{\rm Born}(k,\theta)\right)={2\pi\hbar^2\over m_{\rm red}}~\!a_0~\!
   ,\nonumber
\end{eqnarray}
in which $m_{\rm red}=m_f/2$, because the scattering fermions are identical.
In this way one arrives at
\begin{eqnarray}
  {E_\Omega^{(1)}\over V}
  ={4\pi\hbar^2\over m_f}~\!a_0~\!{p^3_{{\rm F}\uparrow}\over6\pi^2}
  ~\!{p^3_{{\rm F}\downarrow}\over6\pi^2}
  ={\hbar^2\over2m_f}~\!8\pi~\!a_0~\!\rho_\uparrow\rho_\downarrow~\!,
  \label{eqn:E1Omega}
\end{eqnarray}
the result obtained in \cite{LSS}, \cite{Porta} by much more
sophisticated methods as a mathematically rigorous estimate valid
also for arbitrary strengths of the potential
provided the densities $\rho_{\uparrow/\downarrow}$ are small.

\section{Effective Field theory approach}
\label{sec:efftheory}

\noindent Another method of computing the ground state energy, proposed
in \cite{HamFur00}, is based on the effective theory. The most general
Lagrangian density consistent with the Galileo, parity and time-reversal
symmetries the dynamics of the spinor field is assumed to be subjected to,
has the form (spinor indices of field in the brackets are implicitly contracted)
\begin{eqnarray}
  {\cal L}=\psi^\dagger\!\left(\!i\hbar~\!\partial_t
  +{\hbar^2\mbox{\boldmath{$\nabla$}}^2\over2m_f}
  \!\right)\!\psi-{C_0\over2}:\!
  (\psi^\dagger\psi)^2\!:+{C_2\over16}:\![(\psi^\dagger\psi^\dagger)
    (\psi\lrvec{\mbox{\boldmath{$\nabla$}}}^2\!\psi)+{\rm H.c.}]\!:+\dots
  \label{eqn:Leff}
\end{eqnarray}
It consists of infinitely many local operator structures of increasing
dimensions. Their coefficients $C_0$, $C_2$, etc. have
to be determined by comparing the scattering amplitude of two fermions
computed using this effective theory with the one known from the potential
scattering (i.e. the one parametrized by the scattering lengths etc.), or
- see below - by matching onto the ``fundamental'' theory
(\ref{eqn:ModelHamiltonian}).

The first term of the effective interaction $V^{\rm eff}_{\rm int}$ (obtained
as minus the interaction term of the above Lagrangian integrated
over the space) of the effective Hamiltonian\footnote{The effective
theory free Hamiltonian $H_0$ is constructed in the standard
way as the Legendre transform of ${\cal L}_0$ consisting of the terms of
${\cal L}$ bilinear in $\psi^\dagger$ and $\psi$. This obviously gives the
same $H_0$ as in (\ref{eqn:ModelHamiltonian}).}
and, in the finite volume $V$, setting
\begin{eqnarray}
  \psi_\sigma(\mathbf{x})={1\over\sqrt{V}}\sum_{\mathbf{p}}a_{\mathbf{p},\sigma}~\!
    e^{i\mathbf{p}\cdot\mathbf{x}}~\!,\nonumber
\end{eqnarray}
differs from $V_{\rm int}$ of (\ref{eqn:ModelHamiltonian}) only
by the replacement of $\tilde V_{\rm pot}(\mathbf{q})$ by $C_0$ and
obviously reproduces the first order correction $E_\Omega^{(1)}$
to the ground state energy of the system of $N$ fermions, provided
$C_0$ is (in the lowest order) identified with $(4\pi\hbar^2/m_f)a_0$.

In higher orders the local (i.e. singular) nature of the interaction
terms of the Lagrangian density (\ref{eqn:Leff}) results in
short-distance (i.e. ultraviolet) divergences
(evidently absent in the original theory defined by the Hamiltonian
(\ref{eqn:ModelHamiltonian})) in various quantities computed with the
help ot it. These should be regularized and removed by applying the
standard renormalization procedure. Since we are interested here only
in a directly measurable quantity (the ground state energy $E_\Omega$),
renormalization can be straightforwardly carried out by simply computing
(using the same regularization) an appropriate set of observables other
than $E_\Omega$ itself (in this case the scattering lengths $a_\ell$ and
effective radii $r_\ell$ in the expansion of the elastic scattering
amplitude of two fermions) and expressing the computed quantity,
$E_\Omega$, in terms of them.\footnote{Were we, for this or
  another reason, interested in finitness
  of Green's functions, it would be necessary to define them as
  $T$-correlators of appropriately rescaled (misleadingly called
  renormalized in the standard literature on renormalization) Heisenberg
  picture field operators $\psi^{\rm R}=Z_\psi^{-1/2}\psi$ of the
  effective theory. Moreover, in applying the
  renormalization procedure in higher orders it would be much more
  convenient to employ the counterterm technique implemented by
  splitting the ``bare'' couplings (coefficients $C_i$) into
  renormalized ones and the infinite (in the limit of removed
  regularization)
  counterterms $C_i=C_i^{\rm R}+\sum_{k=1}^\infty\delta C^{(k)}_i$
  (and write $Z_\psi=1+\sum_{k=1}^\infty\delta Z_\psi^{(k)}$, in case of
  requiring finiteness of Green's functions; the coefficients $C_i$
  and the Heisenberg picture operators $\psi$ are usually called ``bare''
  in this context)
  and imposing appropriate ``renormalization conditions''. This is
not necessary to the order we will be working.}
Any consistent regularization can be used for this purpose because
when the computed physical quantities (such as $E_\Omega$) are expressed
in terms of other observables ($a_\ell$'s and $r_\ell$'s) they become
independent of it (in the limit of removed regularization).
The most popular in high energy physics computations (in which
preservation of gauge invariance is the main concern) is the dimensional
regularization and it is this one which was used in the seminal paper
\cite{HamFur00}. It however can hardly be given any physical interpretation.
The regularization implemented by cutting off all integrals
over the wave vectors $\mathbf{k}$ at the scale $\Lambda$ is better in
line with the entire effective theory approach which should be viewed
in the wilsonian spirit as follows.\footnote{It is this regularization
   which (in conjunction with the counterterm technique) has been used in the
   recent computation \cite{WeDrSch} of the fourth order corrections to
   $E_\Omega$ of the unpolarized gas of fermions.}
Suppose physical observables are
computed both in the effective theory (\ref{eqn:Leff}) with the cut-off
$\Lambda$ imposed and
directly in the ``fundamental'' theory (\ref{eqn:ModelHamiltonian}).
The (``bare'') coefficients $C_i$ of the Lagrangian (\ref{eqn:Leff})
could be then adjusted to reproduce the results of the
``fundamental'' finite theory (\ref{eqn:ModelHamiltonian}) -
this is the mentioned  matching of the effective theory onto the
fundamental one. This would
make these coefficients explicitly dependent on the cut-off scale
$\Lambda$ and they, of course, would diverge in the (unphysical)
limit $\Lambda\rightarrow\infty$. But obviously all physical quantities
computed in terms of cutoff dependent bare coefficients would be
finite and independent of the scale $\Lambda$, if the dependence of
the bare coefficients $C_i$ on this scale, which can be encoded in the
wilsonian renormalization group equations,\footnote{The counterterm
  technique through the splitting
  $C_i(\Lambda)=C^{\rm R}_i(\mu)+\delta C_i(\Lambda,\mu)$
  allows to introduce an auxiliary scale $\mu$ (introduced
  automatically also in the dimensional regularization) on which
  computed physical quantities do not depend if the explicit dependence
  of renormalized couplings $C^{\rm R}_i(\mu)$ on this scale
  (encoded in the standard renormalization group equations satisfied
  by the renormalized coefficients $C^{\rm R}_i(\mu)$) is taken into account.}
were taken into account. Of course, once observables like $E_\Omega$ and
$a_\ell$, $r_\ell$ are computed in terms of $C_i(\Lambda)$ (obtained
as sketched above), the latter can be eliminated altogether by expressing,
say $E_\Omega$ through $a_\ell$, $r_\ell$'s.\footnote{The proliferation of
  possible operator structures of higher dimensions has the effect that also
  amplitudes of three-body scattering must be used as the input to determine
  all independent coefficients of the effective Lagrangian (\ref{eqn:Leff}).}
The dependence of $E_\Omega$ on $\Lambda$ (originating from cutting off
integrals over the wave vectors) remaining after this operation
is already harmless, and can be removed by formally taking
the limit $\Lambda\rightarrow\infty$ which simplifies
the expressions and is also justified from the practical point
of view, if the (arbitrarily chosen) scale $\Lambda$ of matching
of the fundamental and effective theories is high compared to the
typical physical scale involved in the computed quantity of interest
(in the case of $E_\Omega$ this physical scale is $k_{\rm F}$).
In passing, it is interesting to note that the nonrelativistic theories:
the fundamental one (\ref{eqn:ModelHamiltonian}), 
and the effective one (\ref{eqn:Leff}) provide explicit illustration
of the role of the cutoff scale $\Lambda$ in effective theories on which
an attempt, proposed in \cite{CHLEWMEI}, to resolve the so-called
hierarchy problem of high energy physics was based.
\vskip0.2cm

We now proceed to the determination of $C_0$ and other coefficients
in terms of the scattering lengths $a_\ell$ and effective radii $r_\ell$.
Given (in principle) the spin independent two-body interaction potential
$V_{\rm pot}(\mathbf{r}_1-\mathbf{r}_2)$, the amplitude of elastic scattering
of two nonrelativistic (not necessarily identical at this point)
particles of spins $s_1$ and $s_2$ and masses $m_1$ and $m_2$
is obtained by rewriting the two-body Schr\"odinger equation in terms
of the relative 
$\mathbf{r}=\mathbf{r}_1-\mathbf{r}_2$ and the center of mass 
$\mathbf{R}=(m_1\mathbf{r}_1+m_2\mathbf{r}_2)/(m_1+m_2)$ variables,
separating it by setting
$\Psi(\mathbf{R},\mathbf{r})=\psi(\mathbf{r})\exp(i\mathbf{P}\cdot\mathbf{R}/\hbar)$
and solving the resulting equation (which corresponds to the motion
of a fictitious particle of mass $m_{\rm red}$ with energy $E^\prime$ in the
potential $V_{\rm pot}$)
\begin{eqnarray}
\left(-{\hbar^2\over2m_{\rm red}}~\!\mbox{\boldmath{$\nabla$}}^2_{\mathbf{r}}
+V_{\rm pot}(\mathbf{r})\right)\psi(\mathbf{r})
=E^\prime\Psi(\mathbf{r})~\!,\label{eqn:SchrEq}
\end{eqnarray}
in which $m_{\rm red}=m_1m_2/(m_1+m_2)$ and 
$E^\prime=E-\hbar^2\mathbf{P}^2/2M$, in the center of mass system
(CMS) that is, setting $\mathbf{P}=\mathbf{0}$. In this frame
the momenta of the two particles are $\hbar\mathbf{k}$
and $-\hbar\mathbf{k}$, respectively, and the energy
$E^\prime=E$ of the fictitious particle
is the total energy of the two colliding particles:
\begin{eqnarray}
  E={1\over2}m_1\mathbf{v}_1^2+{1\over2}m_2\mathbf{v}_2^2
  ={\hbar^2\over2}\left(m_1~\!{\mathbf{k}^2\over m_1^2}+m_2~\!{\mathbf{k}^2\over m_2^2}
  \right)={\hbar^2\mathbf{k}^2\over2m_{\rm red}}~\!.\nonumber
\end{eqnarray}
It follows that the parameter $\mathbf{k}$ playing the role of the wave
vector of the fictitious particle should be identified
with the wave vector of one of the scattering particles (the true relative
momentum of the two scattering particles is $2\mathbf{k}$).
As long as particles are distinguishable (and the interaction
potential is spin independent), individual spins of the two
particles and their total spin $S$ are preserved and do not play
any role: the scattering amplitude $f(k,\theta)$ is defined in terms
of the asymptotic form ($k\equiv|\mathbf{k}|$)
\begin{eqnarray}
  \psi^{(+)}_{\mathbf{k}}(\mathbf{r})=e^{i\mathbf{k}\cdot\mathbf{r}}
  +{f(k,\theta)\over r}~\!e^{ikr} ~\!,\nonumber
\end{eqnarray}
of the ({\it in}-state) solution of the Schr\"odinger equation
(\ref{eqn:SchrEq}) with $E^\prime=\hbar^2\mathbf{k}^2/2m_{\rm red}$ and 
the differential scattering cross section $d\sigma/d\Omega$
is simply given by $|f(k,\theta)|^2$.

If the two particles are identical and both have spin $s$ (integer or
half-integer), the complete wave function of the system written as
a product $\Psi(\mathbf{r}_1,\mathbf{r}_2)\chi(\sigma_1,\sigma_2)$,
where $\sigma_{1,2}=-s,\dots,+s$ must be either totally symmetric or
totally antisymmetric depending on whether $s$ is integer or
half-integer. If the interaction is spin-independent, one can chose
$\chi(\sigma_1,\sigma_2)$ to be symmetric or antisymmetric.
$\psi(\mathbf{r})$ must, therefore, also be constructed as
symmetric $\psi(-\mathbf{r})=\psi(\mathbf{r})$ or antisymetric
and the simple rule \cite{LL} is then that if the total spin $S$
of the two particles is even, $\psi(\mathbf{r})$ must be symmetric,
and antisymmetric, when the total spin $S$ is odd.
Thus, if  $s={1\over2}$ but $S=0$, the scattering amplitude of the two
indistinguishable spin $1/2$ fermions is given
by $f(k,\theta)+f(k,\pi-\theta)$, while if $S=1$, it is 
 $f(k,\theta)-f(k,\pi-\theta)$.

In the formalism of second quantization one computes the $S$-matrix
element\footnote{The formula (\ref{eqn:Smatrix}) is based on the very
  strong assumption that there is a well defined, strict one-to-one
  correspondence between the {\it in} and {\it out} eigenstates of
  the complete Hamiltonian and the eigenstates
  of $H_0$. In most relativistic field theories, except for the simplest ones,
  like $\varphi^4$ models and quantum electrodynamics (which probably
  explains its ubiquity in elementary textbooks without any qualifications)
  these assumptions cannot be satisfied. Even in cases the required
  one-to-one correspondence 
  can be enforced, the use of the formula (\ref{eqn:Smatrix}) necessarily
  entails working in the so-called On-Shell renormalization scheme,
  that is with physical mass $m_{\rm ph}$ in the free propagator
  and with physically normalized Heisenberg picture field operators.
  In more complicated cases $S$-matrix elements can only be obtained by
  applying the LSZ prescription to appropriate Green's functions.
  In the nonrelativistic theory considered here the assumptions are
  satisfied precisely owing to the form of the interaction
  (its normal ordering with respect to the vector $|{\rm void}\rangle$) and
  to the absence of antiparticles
  which together preclude any genuine self-interaction of the particles.}
\cite{Weinb}
\begin{eqnarray}
  S_{\beta\alpha}=\langle\beta_0|{\rm T}\exp\!\left(-{i\over\hbar}\!\int_{-\infty}^\infty
  \!dt~\!V_{\rm int}^I(t)\right)\!|\alpha_0\rangle~\!,\label{eqn:Smatrix}
\end{eqnarray}
in which as $|\alpha_0\rangle$ and
$|\beta_0\rangle$ are the states of two fermions:
\begin{eqnarray}
  |\alpha_0\rangle=a^\dagger_{\sigma_2}(\mathbf{k}_2)~\!a^\dagger_{\sigma_1}(\mathbf{k}_1)
  |{\rm void}\rangle~\!,\nonumber\\
  |\beta_0\rangle=a^\dagger_{\sigma_2^\prime}(\mathbf{k}^\prime_2)~\!
  a^\dagger_{\sigma_1^\prime}(\mathbf{k}^\prime_1)|{\rm void}\rangle~\!,~\!\nonumber
\end{eqnarray}
and $V_{\rm int}^I(t)$ is the interaction picture counterpart of the
interaction term of the Hamiltonian (\ref{eqn:ModelHamiltonian})
written in the position representation (in the normalization in the
continuum, as requires the application to the scattering theory)
\begin{eqnarray}
  V_{\rm int}={1\over2}\!\int\!d^3\mathbf{x}\!\int\!d^3\mathbf{y}~\!
  \psi^\dagger_\alpha(\mathbf{x})\psi^\dagger_\beta(\mathbf{y})~\!
  V_{\rm pot}(\mathbf{x}-\mathbf{y})~\!\psi_\beta(\mathbf{y})\psi_\alpha(\mathbf{x})~\!.
  \nonumber
\end{eqnarray}
that is, expressed in terms of the interaction picture field operators
\begin{eqnarray}
  \psi^I_\sigma(t,\mathbf{x})=e^{iH_0t/\hbar}~\!\psi_\sigma(\mathbf{x})~\!e^{-iH_0t/\hbar}
  =\int\!{d^3\mathbf{k}\over(2\pi)^3}~\!
  e^{-i\omega_{\mathbf{k}}t+i\mathbf{k}\cdot\mathbf{x}}~\!a_\sigma(\mathbf{k})~\!,
  \nonumber
\end{eqnarray}
in which $\omega_{\mathbf{k}}\equiv\hbar\mathbf{k}^2/2m_f$ (the notation
$k^0=\omega_{\mathbf{k}}$ will also be employed below).

Evaluated using the standard methods (Wick theorem, etc.) the
$S$-matrix element takes, in the case of spin independent interaction,
the general form
\begin{eqnarray}
  S_{\beta\alpha}=\langle\beta_0|\alpha_0\rangle-{i\over\hbar}~\!(2\pi)^4
  \delta(\omega_{\mathbf{k}_1^\prime}+\omega_{\mathbf{k}_2^\prime}
  -\omega_{\mathbf{k}_1}-\omega_{\mathbf{k}_2})~\!
  \delta^{(3)}(\mathbf{k}_1^\prime+\mathbf{k}_2^\prime-\mathbf{k}_1-\mathbf{k}_2)
  \phantom{aaaaa}\nonumber\\
  \times\left[{\cal A}\!\left((\mathbf{k}_1-\mathbf{k}^\prime_1)^2\right)
    \delta_{\sigma_1^\prime\sigma_1}
    \delta_{\sigma_2^\prime\sigma_2}
    -{\cal A}\!\left((\mathbf{k}_1-\mathbf{k}^\prime_2)^2\right)
    \delta_{\sigma_1^\prime\sigma_2}
    \delta_{\sigma_2^\prime\sigma_1}\right].
\end{eqnarray}
Since $(\mathbf{k}_1-\mathbf{k}_1^\prime)^2=4|\mathbf{k}|^2\sin^2(\theta/2)$,
$(\mathbf{k}_1-\mathbf{k}_2^\prime)^2=4|\mathbf{k}|^2\sin((\pi-\theta)/2)$,
the two terms in the bracket must be proportional to the nonrelativistic
amplitudes $f(k,\theta)$ and $f(k,\pi-\theta)$. It is then easy to
check\footnote{The simplest way is to take a concrete potential, e.g.
  of the Yukawa type,
  $V_{\rm pot}(\mathbf{x})=(g^2/|\mathbf{x}|)\exp(-M_\phi c|\mathbf{x}|/\hbar)$,
  and  to compare the first order term in ${\cal A}$ obtained from
  (\ref{eqn:Smatrix}) with the Born formula for $f(k,\theta)$, remembering
  that in the latter $m_{\rm red}=m_f/2$. The Landau-Lifschitz rules
  can also be checked by constructing the $|\alpha_0\rangle$ and $|\beta_0\rangle$
states with definite values of the total spin $S$.}
that the proper identification reads
\begin{eqnarray}
  -{m_f\over4\pi\hbar^2}~\!{\cal A}(k,\theta)=f(k,\theta)~\!.\label{eqn:Rule}
\end{eqnarray}

In general, the scattering amplitude $f(k,\theta)$ can be expressed
through the partial wave shifts as
\begin{eqnarray}
  f(k,\theta)={1\over k}\sum_{l=0}^\infty(2l+1)~\!e^{i\delta_l(k)}\sin\delta_l(k)~\!
  P_l(\cos\theta)
  =\sum_{l=0}^\infty(2l+1)~\!{1\over k~\!{\rm cot}\delta_l-ik}~\!
  P_l(\cos\theta)~\!,\nonumber
\end{eqnarray}
and the scattering lengths $a_l$ and the effective ranges $r_l$
are defined as the coefficients in the small $k$ expansions
\begin{eqnarray}
  k~\!{\rm cot}\delta_0=-{1\over a_0}+{1\over2}~\!r_0k^2+\dots,\phantom{aaa}
  k~\!{\rm cot}\delta_1=-{3\over k^2a_1^3}+\dots\nonumber
\end{eqnarray}
Therefore,  the small $k$ expansion of the scattering amplitude reads
\begin{eqnarray}
  f(k,\theta)=-a_0\left[1-ia_0k+\left({1\over2}~\!a_0r_0-a_0^2\right)k^2
    +\dots\right]-a_1^3k^2\cos\theta+\dots,\label{eqn:fScattExp}
\end{eqnarray}

Evaluating the formula (\ref{eqn:Smatrix}) for
\begin{eqnarray}
  |\alpha_0\rangle=a^\dagger_\downarrow(\mathbf{k}_2)
  a^\dagger_\uparrow(\mathbf{k}_1)|{\rm void}\rangle~\!,
  \nonumber\\
  |\beta_0\rangle=a^\dagger_\downarrow(\mathbf{k}^\prime_2)
  a^\dagger_\uparrow(\mathbf{k}^\prime_1)|{\rm void}\rangle~\!,\nonumber
\end{eqnarray}
(in order to extract directly one part of the scattering amplitude)
one easily gets
\begin{eqnarray}
  S_{\beta\alpha}=\langle\beta_0|\alpha_0\rangle-{i\over\hbar}~\!C_0
  \times(2\pi)^4\delta^{(4)}(k_1^\prime+k_2^\prime-k_1-k_2)~\!,\nonumber
\end{eqnarray}
(the wave vectors $k_i$'s are treated as four-vectors: $k^0=\omega_{\mathbf{k}}$). 
Thus ${\cal A}=C_0$ simply. Using the rule (\ref{eqn:Rule})
one obtains that in the first order  $C_0=(4\pi\hbar^2/m_f)a_0$.

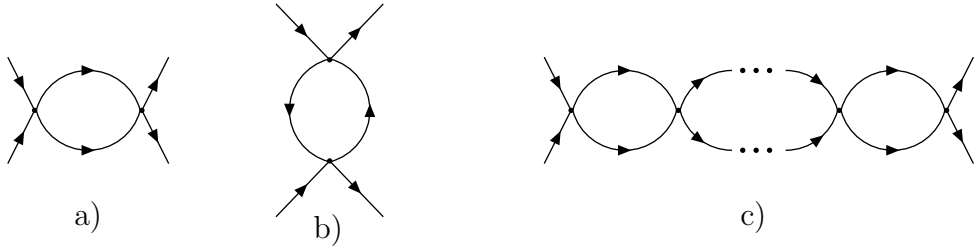
\begin{figure}[]
\begin{center}
\begin{picture}(400,80)(5,0)
\ArrowArc(30,45)(20,195,345)
\ArrowArcn(30,35)(20,165,15)
\Vertex(10,40){1}
\Vertex(50,40){1}
\ArrowLine(0,20)(10,40)
\ArrowLine(0,60)(10,40)
\ArrowLine(50,40)(60,20)
\ArrowLine(50,40)(60,60)
\Text(30,0)[]{a)}
\ArrowArc(115,40)(20,285,75)
\ArrowArc(125,40)(20,105,255)
\Vertex(120,21){1}
\Vertex(120,59){1}
\ArrowLine(100,0)(120,21)
\ArrowLine(120,21)(140,0)
\ArrowLine(100,80)(120,59)
\ArrowLine(120,59)(140,80)
\Text(120,-5)[]{b)}
\ArrowArc(230,45)(20,195,345)
\ArrowArcn(230,35)(20,165,15)
\ArrowArc(270,45)(20,195,270)
\ArrowArcn(270,35)(20,165,90)
\Vertex(274,55){1}
\Vertex(279,55){1}
\Vertex(284,55){1}
\Vertex(274,25){1}
\Vertex(279,25){1}
\Vertex(284,25){1}
\Text(279,0)[]{c)}
\ArrowArc(290,45)(20,270,345)
\ArrowArcn(290,35)(20,90,15)
\ArrowArc(330,45)(20,195,345)
\ArrowArcn(330,35)(20,165,15)
\Vertex(210,40){1}
\Vertex(250,40){1}
\Vertex(310,40){1}
\Vertex(350,40){1}
\ArrowLine(200,20)(210,40)
\ArrowLine(200,60)(210,40)
\ArrowLine(350,40)(360,20)
\ArrowLine(350,40)(360,60)
\end{picture}
\end{center}
\caption{Two one-loop (second order)
  diagrams potentially contributing to the elastic scattering
  amplitude of two fermions: diagram a) is nonzero, diagram b) vanishes.
  The ``sausage''-type diagrams c) arising in higher orders.
  Time flows from the left to the right.}
\label{fig:ScatteringOneLoop}
\end{figure}

The next order contribution to $S_{\beta\alpha}$ is (we omit the superscript $I$
- interaction picture - on operators; $d^4x\equiv dt~\!d\mathbf{x}$)
\begin{eqnarray}
  {1\over2!}\left({C_0\over2i\hbar}\right)^2\!\int\!d^4x\!\int\!d^4y~\!
  \langle a_{1^\prime}a_{2^\prime}~\!T\!\left[:\!
    (\psi^\dagger_\sigma\psi_\sigma)^2(x)\!: ~\!\!\!:\!
    (\psi^\dagger_{\sigma^\prime}\psi_{\sigma^\prime})^2(y)\!:\right]
a^\dagger_2 a^\dagger_1\rangle~\!.\nonumber
\end{eqnarray}
Application of the Wick theorem to the operators under the symbol of the
chronological ordering leads to several different terms of which one needs
those which have two $\psi$ and two $\psi^\dagger$ operators unpaired (they
are needed to remove the operators building the initial and final states).
We will consider first the term which corresponds to the diagram of
Fig.~\ref{fig:ScatteringOneLoop}a. Its contribution is
\begin{eqnarray}
  \left({C_0\over i\hbar}\right)^2\!\int\!d^4x\!\int\!d^4y~\!e^{ik_2^\prime x}e^{ik_1^\prime x}~\!
  iG^{(0)}(x-y)~\!iG^{(0)}(x-y)~\!e^{-ik_2y}~\!e^{-ik_1y}~\!,\nonumber
\end{eqnarray}
The propagator ($iG^{(0)}_{\alpha\beta}(x-y)=\delta_{\alpha\beta}iG^{(0)}(x-y)$)
is given by 
\begin{eqnarray}
  iG^{(0)}_{\alpha\beta}(x-y)=\langle{\rm void}|T[\psi_\alpha(x)\psi_\beta^\dagger(y)]
  |{\rm void}\rangle\phantom{aaaaaaaaaaaaaaaaaaaaaa}~\!\nonumber\\
  =\delta_{\alpha\beta}
  \int\!{d^3\mathbf{q}\over(2\pi)^3}\int_{-\infty}^\infty\!{d\omega\over2\pi}~\!
  {i\over\omega-\omega_{\mathbf{q}}+i0}~\!
  e^{-i\omega(x^0-y^0)+i\mathbf{q}\cdot(\mathbf{x}-\mathbf{y})}~\!.\nonumber
\end{eqnarray}
Using this form of the propagator, the integrals over $d^4x$ and $d^4y$ can be
taken and one gets $(2\pi)^4\delta^{(4)}(k_1^\prime+k_2^\prime-k_1-k_2)$ times
\begin{eqnarray}
  \left({C_0\over i\hbar}\right)^2\!
  \int\!{d^3\mathbf{q}\over(2\pi)^3}\int_{-\infty}^\infty\!{d\omega\over2\pi}~\!
  {i^2\over[\omega-\omega_{\mathbf{q}}+i0]
  [-\omega+k_1^0+k_2^0-\omega_{\mathbf{k}_1+\mathbf{k}_2-\mathbf{q}}+i0]}~\!.
  \nonumber
\end{eqnarray}
The integral over $d\omega$ is nonzero because the two simple poles of
the integrand are located one above and one below the real $\omega$
axis. It should be remarked here that the right diagram shown in Fig.
\ref{fig:ScatteringOneLoop}b gives zero because it leads to a similar
integral in which, however, both poles are below the real $\omega$ axis;
the integral over $d\omega$ then gives zero.\footnote{Other diagrams
  which one could draw having two incoming, two outgoing lines and two
  vertices are in fact absent because the interaction $V_{\rm int}$ is
  normally ordered with respect to the $|{\rm void}\rangle$ state.}
This simplification arises because of the absence of antiparticle in the
nonrelativistic theory. Picking up the pole below the real axis to
perform the integral over $d\omega$
and then working out the denominator ($\omega_{\mathbf{q}}=\hbar\mathbf{q}^2/2m_f$,
$k_1^0=\hbar\mathbf{k}_1^2/2m_f$, etc.) one obtains
\begin{eqnarray}
  \left({C_0\over i\hbar}\right)^2\!\int\!{d^3\mathbf{q}\over(2\pi)^3}~\!
  {-i~\!m_f/\hbar\over\mathbf{q}^2-\mathbf{q}\!\cdot\!(\mathbf{k}_1+\mathbf{k}_2)+
  \mathbf{k}_1\!\cdot\!\mathbf{k}_2-i0}~\!.\nonumber
\end{eqnarray}
Instead of evaluating this integral keeping the momenta $\mathbf{k}_1$
and $\mathbf{k}_2$ arbitrary, one can take advantage of the fact that
the scattering amplitude is computed in the CMS in which
$\mathbf{k}_1+\mathbf{k}_2=\mathbf{0}$ and
$\mathbf{k}_1\cdot\mathbf{k}_2=-\mathbf{k}^2$. This reduces the integral to
the form (7) in \cite{HamFur00}:
\begin{eqnarray}
  I_0=\int\!{d^3\mathbf{q}\over(2\pi)^3}~\!
  {1\over\mathbf{q}^2-\mathbf{k}^2-i0}~\!.\label{eqn:I0integralDef}
\end{eqnarray}
It is not difficult to see that there is a whole class of diagrams
shown in Fig.~\ref{fig:ScatteringOneLoop}c
representing the elastic scattering which is generated by the interaction
term proportional to $C_0$. Their contribution to $S_{\beta\alpha}$ can
immediately be written down:
\begin{eqnarray}
  (2\pi)^4\delta^{(4)}(k_1^\prime+k^\prime_2-k_1-k_2)\!
  \left\{{C_0\over i\hbar}+\left({C_0\over i\hbar}\right)^2\!{m_f\over i\hbar}~\!I_0
  +\left({C_0\over i\hbar}\right)^3\!
  \left({m_f\over i\hbar}~\!I_0\right)^2+\dots\right\}.\nonumber
\end{eqnarray}
This gives the scattering amplitude ${\cal A}=i\hbar\{\dots\}$ and,
according to the rule (\ref{eqn:Rule}) the corresponding
contribution to the scattering amplitude  $f(k,\theta)$ is
\begin{eqnarray}
  f(k,\theta)=-{m_f\over4\pi\hbar^2}~\!{\cal A}=
  -{m_f\over4\pi\hbar^2}~\!C_0\left\{1
  +\left({C_0\over i\hbar}\right)\left({m_f\over i\hbar}~\!I_0\right)
  +\left({C_0\over i\hbar}\right)^2
  \left({m_f\over i\hbar}~\!I_0\right)^2+\dots\right\}.\label{eqn:fAmp}
\end{eqnarray}
This amplitude, supplemented with terms which come from tree level diagrams
generated by the interactions proportional to the coefficients $C_2$ and
$C_2^\prime$, and with yet higher orders from loop diagrams involving all
other vertices present in the effective Lagrangian should be matched
onto the expansion (\ref{eqn:fScattExp}).
\vskip0.2cm

The integral $I_0$ is divergent and requires regularization.
Instead of using for this purpose the dimensional regularization as
in \cite{HamFur00}, we regularize it by imposing the UV cut-off
$\Lambda$ on $q=|\mathbf{q}|$. 
\begin{eqnarray}
  I_0(k,\Lambda)={1\over2\pi^2}\!\int_0^\Lambda\!{dq~\!q^2\over q^2-k^2-i0}
  ={1\over4\pi^2}\!\int_0^\Lambda\!dq~\!q\left[{1\over q-k-i0}
    +{1\over q+k+i0}\right]. \nonumber
\end{eqnarray}
Since the integral runs from zero, the pole is only in the first part of
the integrand, and, upon using the Sochocki formula ($P$ stands for
principal value), one obtains
\begin{eqnarray}
  I_0(k,\Lambda)
  ={i\over4\pi}~\!k+{1\over2\pi^2}~\!\Lambda-{1\over2\pi^2}~\!{k^2\over\Lambda}+\dots
 \label{eqn:I0integralResult}
\end{eqnarray}
Inserting this into the formula (\ref{eqn:fAmp}) matched onto the expansion
(\ref{eqn:fScattExp}) and 
solving for (real) $C_0$  setting
$C_0=(4\pi\hbar^2/m_f)a_0(1+\Delta/a_0)$
one obtains 
\begin{eqnarray}
  C_0={4\pi\hbar^2\over m_f}~\!a_0\left(1+{2\over\pi}~\!a_0\Lambda +\dots\right).
  \label{eqn:C0determined}
\end{eqnarray}
The procedure leading to $C_0$ is more complicated when the cut-off
regularization is used and the resulting $C_0$ is given in the form of
an infinite power series in $a_0\Lambda$ (this is why the dimensional
regularization is more convenient) but, as will be seen, allows to
better control cancellation of divergences in physical quantities.

\section{Corrections to the ground state energy}
\label{sec:Ecorrections}

We now compute the correction $E_\Omega^{(2)}$ to the
ground state energy diagrammatically, treating the terms of the  Lagrangian
(\ref{eqn:Leff}) as the interaction vertices. The calculation closely
parallels that of \cite{HamFur00} except that we do not  assume that
the numbers $N_\uparrow$ and  $N_\downarrow$ of spin up and spin down fermions
are equal. The basic formula employed for this purpose reads\footnote{The
  symbol T of the chronological ordering should not be confused with
  $T$ denoting time.}
\begin{eqnarray}
  \lim_{T\rightarrow\infty}\exp(-iT(E_\Omega-E_{\Omega_0})/\hbar)=
  \lim_{T\rightarrow\infty}\langle\Omega_0
  |{\rm T}\exp\!\left(-{i\over\hbar}\!\int_{-T/2}^{T/2}\!dt~\!
  V^I_{\rm int}(t)\right)\!|\Omega_0\rangle~\!.\label{eqn:basicFormula}
\end{eqnarray}
In other words, $-iT(E_\Omega-E_{\Omega_0})/\hbar$ is (in the limit
$T\rightarrow\infty$,
$V\rightarrow\infty$) given by $(2\pi)^4\delta^{(4)}(0)$ times the sum of
the momentum space connected vacuum diagrams (diagrams without external
lines). The factor $(2\pi)^4\delta^{(4)}(0)$ arising in evaluating diagrams
in position space (expressing the overall four-momentum conservation)
is interpreted as $VT$. It follows that $i\hbar$ times the expression
arising from summing the momentum space connected vacuum diagrams is
just $(E_\Omega-E_{\Omega_0})/V$. 

As explained in \cite{HamFur00}, to compute the order $(k_{\rm F}a_0)^2$
correction to the ground state energy only the interaction term
proportional to $C_0$ of (\ref{eqn:Leff}) is needed. It simplifies
considerably if there are only two possible spin projections, because
$\psi_\alpha^\dagger\psi_\alpha^\dagger=\psi_\alpha\psi_\alpha=0$ (one can treat
fields as anticommuting), and reads\footnote{In what follows we will
denote by $+$ ($-$) quantities and operators pertaining to the spin
projection of larger (smaller) density; thus we will use the Fermi
momenta $p_{{\rm F}+}$ and $p_{{\rm F}-}$ ($p_{{\rm F}+}\geq p_{{\rm F}-}$)
understanding that $p_{{\rm F}+}=p_{{\rm F}\uparrow}$ when
$p_{{\rm F}\uparrow}\geq p_{{\rm F}\downarrow}$.}
\begin{eqnarray}
  V_{\rm int}=C_0\!\int\!d^3\mathbf{x}:\!(\psi^\dagger_+\psi_+)
  (\psi^\dagger_-\psi_-)\!:~\!=C_0\!\int\!d^3\mathbf{x}:\!(\psi^\dagger_-\psi_-)
  (\psi^\dagger_+\psi_+)\!:~\!.\label{eqn:psiplusminus}
\end{eqnarray}
In evaluating the right hand side of the formula (\ref{eqn:basicFormula})
expanding the exponents and using the Wick theorem needed are the
propagators (see e.g. \cite{FetWal})
\begin{eqnarray}
  iG^{(0)}_\pm(t^\prime,\mathbf{x}^\prime,t,\mathbf{x})
  \equiv\langle\Omega_0|{\rm T}\psi^I_\pm(t^\prime,\mathbf{x}^\prime)
  (\psi^I_\pm(t,\mathbf{x}))^\dagger|\Omega_0\rangle
  \phantom{aaaaaaaaaaaaaa}\!\nonumber\\
  =\int_{-\infty}^\infty\!{d\omega\over2\pi}\!\int\!{d^3\mathbf{k}\over(2\pi)^3}
  ~\!e^{-i\omega(t^\prime-t)}~\!e^{i\mathbf{k}\cdot(\mathbf{x}^\prime-\mathbf{x})}
  ~\!i\tilde G^{(0)}_\pm(\omega,\mathbf{k})~\!,\nonumber
\end{eqnarray}
(all other Wick contractions vanish) in which
\begin{eqnarray}
  i\tilde G^{(0)}_\pm(\omega,\mathbf{k})=i\left[{\theta(|\mathbf{k}|-p_{{\rm F}\pm})\over
      \omega-\omega_{\mathbf{k}}+i0}+{\theta(p_{{\rm F}\pm}-|\mathbf{k}|)\over
      \omega-\omega_{\mathbf{k}}-i0}\right],\label{eqn:MomSpaceProp}
\end{eqnarray}
where $\omega_{\mathbf{k}}=\hbar\mathbf{k}^2/2m_f$. Analytical expressions
are ascribed to individual diagrams according to the standard rules of
quantum field theory; to account for normal ordered form of the interaction
one has only to add the rule \cite{FetWal} that if a line originates from
and ends up in one and the same vertex, the propagator
(\ref{eqn:MomSpaceProp}) corresponding to this line
has to be multiplied by $e^{i\omega\eta}$ with the
limit $\eta\rightarrow0^+$ taken at the end.

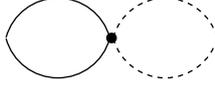
\begin{figure}[]
\begin{center}
\begin{picture}(80,40)(5,0)
\CArc(20,35)(20,195,345)
\CArc(20,25)(20,15,165)
\DashCArc(60,35)(20,195,345){2}
\DashCArc(60,25)(20,15,165){2}
\Vertex(40,30){2}
%
\end{picture}
\end{center}
\caption{The effective theory connected vacuum diagram of order $C_0$
  reproducing the first order correction $E_\Omega^{(1)}$.
  Solid and dashed lines represent propagators of
  fermions with opposite spin projections.}
\label{fig:VacuumDiagrSpinOneHalf}
\end{figure}

In the first order in $C_0$ there is only one connected vacuum
graph shown in Figure \ref{fig:VacuumDiagrSpinOneHalf}. 
Applying the Feynman rules one easily obtains
\begin{eqnarray}
  T E^{(1)}_\Omega
  =C_0~\!VT~\!iG_+(0)~\!iG_-(0)=C_0~\!VT~\!{p^3_{\rm F-}\over6\pi^2}
  ~\!{p^3_{\rm F+}\over6\pi^2}~\!,\nonumber
\end{eqnarray}
and recovers, after using (up to the first order in $a_0$) the result
(\ref{eqn:C0determined}), the formula (\ref{eqn:E1Omega}).
\vskip0.2cm

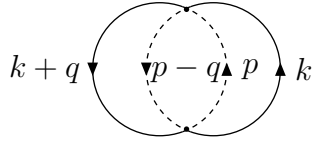
\begin{figure}[]
\begin{center}
\begin{picture}(80,40)(5,0)
\ArrowArc(30,20)(25,70,290)
\DashArrowArc(30,20)(25,290,70){2}
\DashArrowArc(50,20)(25,110,250){2}
\ArrowArc(50,20)(25,250,110)
\Vertex(40,-2.5){1}
\Vertex(40,42.5){1}
\Text(84,20)[]{$k$}
\Text(64,20)[]{$p$}
\Text(-13,20)[]{$k+q$}
\Text(40,20)[]{$p-q$}
\end{picture}
\end{center}
\caption{The only nonvanishing three loop connected vacuum diagram
  contributing to the ground state energy of the diluted gas of
  spin ${1\over2}$ fermions. The solid and dashed lines correspond to
  propagators of fermions having opposite spin projections. The
  two kinds of propagators differ by the values of the Fermi momenta:
  we assume that $p_{{\rm F}+}\geq p_{{\rm F}-}$.}
\label{fig:3LoopVacuumDifferent}
\end{figure}

Several connected vacuum diagrams of the next order can be drawn but, as explained
in \cite{HamFur00}, nonzero
is only the one shown in Figure \ref{fig:3LoopVacuumDifferent}.
The contribution of this diagram is given by 
\begin{eqnarray}
  {E^{(2)}_\Omega\over V}=-{i\over2!}~\!{C_0^2\over\hbar}
  \int\!{d^4q\over(2\pi)^4}\!\int\!{d^4p\over(2\pi)^4}\!
  \int\!{d^4k\over(2\pi)^4}~\!i\tilde G_-(p)~\!i\tilde G_-(p-q)~\!
  i\tilde G_+(k)~\!i\tilde G_+(k+q)\nonumber\\
=-{i\over2!}~\!{C_0^2\over\hbar}
  \int\!{d^3\mathbf{q}\over(2\pi)^3}\!\int\!{d^3\mathbf{p}\over(2\pi)^3}\!
  \int\!{d^3\mathbf{k}\over(2\pi)^3}\!\int_{-\infty}^\infty{dq_0\over2\pi}
  \left[\matrix{\phantom{a}\cr\phantom{a}}\!\!\!\!\!\dots\right],
  \phantom{aaaaaaaaaaaaaaaaaa}\nonumber
\end{eqnarray}
where $[\dots]$ stands for the product of two integrals over frequencies 
\begin{eqnarray}  
  \int_{-\infty}^\infty{dk_0\over2\pi}\left[{\theta(|\mathbf{k}|-p_{\rm F+})
      \over k_0-\omega_{\mathbf{k}}+i0}+{\theta(p_{\rm F+}-|\mathbf{k}|)
      \over k_0-\omega_{\mathbf{k}}-i0}\right]\!
  \left[{\theta(|\mathbf{k}+\mathbf{q}|-p_{\rm F+})
      \over k_0+q_0-\omega_{\mathbf{k}+\mathbf{q}}+i0}
    +{\theta(p_{\rm F+}-|\mathbf{k}+\mathbf{q}|)
      \over k_0+q_0-\omega_{\mathbf{k}+\mathbf{q}}-i0}\right]~\!\nonumber\\
   \int_{-\infty}^\infty{dp_0\over2\pi}\left[{\theta(|\mathbf{p}|-p_{\rm F-})
      \over p_0-\omega_{\mathbf{p}}+i0}+{\theta(p_{\rm F-}-|\mathbf{p}|)
      \over p_0-\omega_{\mathbf{p}}-i0}\right]\!
  \left[{\theta(|\mathbf{p}-\mathbf{q}|-p_{\rm F-})
      \over p_0-q_0-\omega_{\mathbf{p}-\mathbf{q}}+i0}
    +{\theta(p_{\rm F-}-|\mathbf{p}-\mathbf{q}|)
      \over p_0-q_0-\omega_{\mathbf{p}-\mathbf{q}}-i0}\right].\nonumber
\end{eqnarray}

Each of these integrals gives only two terms (not four) because 
the two poles of the integrands must lie one above and the other
one below the real axis - the integrands with both poles above
or both poles below the axis give zero. Performing next the integrations
over $dk_0$, $dp_0$ and then over $dq_0$ one gets
\begin{eqnarray}
  \int_{-\infty}^\infty{dq_0\over2\pi}
  \left[\matrix{\phantom{a}\cr\phantom{a}}\!\!\!\!\!\dots\right]
  =  i~\!{\theta(p_{\rm F+}-|\mathbf{k}|)~\!
    \theta(p_{\rm F-}-|\mathbf{p}|)~\!
    \theta(|\mathbf{k}+\mathbf{q}|-p_{\rm F+})~\!
    \theta(|\mathbf{p}-\mathbf{q}|-p_{\rm F-})\over
    \omega_{\mathbf{k}}+\omega_{\mathbf{p}}-\omega_{\mathbf{k}+\mathbf{q}}
    -\omega_{\mathbf{p}-\mathbf{q}}+i0}\phantom{a}\nonumber\\
  -i~\!{\theta(|\mathbf{k}|-p_{\rm F+})~\!
    \theta(|\mathbf{p}|-p_{\rm F-})~\!
    \theta(p_{\rm F+}-|\mathbf{k}+\mathbf{q}|)~\!
    \theta(p_{\rm F-}-|\mathbf{p}-\mathbf{q}|)\over
    \omega_{\mathbf{k}}+\omega_{\mathbf{p}}-\omega_{\mathbf{k}+\mathbf{q}}
    -\omega_{\mathbf{p}-\mathbf{q}}-i0}~\!.\nonumber
\end{eqnarray}
Making then the substitutions $\mathbf{k}+\mathbf{q}=-\mathbf{k}^\prime$,
$\mathbf{p}-\mathbf{q}=-\mathbf{p}^\prime$ in the second integral one
finds that the two terms above give equal contributions. Thus,
\begin{eqnarray}
  {E^{(2)}_\Omega\over V}={C_0^2\over\hbar}\!\!
  \int\!{d^3\mathbf{q}\over(2\pi)^3}\!\!\int\!{d^3\mathbf{p}\over(2\pi)^3}\!\!
  \int\!{d^3\mathbf{k}\over(2\pi)^3}
{\theta(p_{\rm F+}-|\mathbf{k}|)~\!
    \theta(p_{\rm F-}-|\mathbf{p}|)~\!
    \theta(|\mathbf{k}+\mathbf{q}|-p_{\rm F+})~\!
    \theta(|\mathbf{p}-\mathbf{q}|-p_{\rm F-})\over
    \omega_{\mathbf{k}}+\omega_{\mathbf{p}}-\omega_{\mathbf{k}+\mathbf{q}}
    -\omega_{\mathbf{p}-\mathbf{q}}+i0}~\!.
  \nonumber
\end{eqnarray}
The last step \cite{HamFur00} is to pass to the integrations over the
variables $\mathbf{s}$, $\mathbf{t}$ and $\mathbf{u}$ defined by the
relations (the Jacobian $J=8$)
\begin{eqnarray}
  \mathbf{k}=(\mathbf{s}-\mathbf{t})~\!,\phantom{aaa}
  \mathbf{p}=(\mathbf{s}+\mathbf{t})~\!,\phantom{aaa}
  \mathbf{q}=(\mathbf{t}-\mathbf{u})~\!.\nonumber
\end{eqnarray}
The denominator of the integrand then becomes equal
$(\mathbf{t}^2-\mathbf{u}^2)/m_f$. Defining 
\begin{eqnarray}
  I=\int\!d^3\mathbf{s}\!\int\!d^3\mathbf{t}\!
  \int\!d^3\mathbf{u}~\!
  {\theta(p_{\rm F-}-|\mathbf{t}+\mathbf{s}|)~\!\theta(p_{\rm F+}-|\mathbf{t}-\mathbf{s}|)~\!
  \theta(|\mathbf{u}+\mathbf{s}|-p_{\rm F-})~\!\theta(|\mathbf{u}-\mathbf{s}|-p_{\rm F+})
\over\mathbf{t}^2-\mathbf{u}^2+i0}~\!,\nonumber
\end{eqnarray}
one can write the combined first and second order contributions in the form
\begin{eqnarray}
  {E^{(1)}_\Omega+E^{(2)}_\Omega\over V}=C_0~\!{p^3_{\rm F-}p^3_{\rm F+}\over36\pi^4}
  +{8C_0^2\over\hbar^2}~\!m_f~\!{I\over(2\pi)^9}~\!.\nonumber
\end{eqnarray}
After using (\ref{eqn:C0determined}), i.e. replacing $C_0$ by
$(4\pi\hbar^2/m_f)a_0(1+2a_0\Lambda/\pi)$, one gets
\begin{eqnarray}
  {E^{(1)}_\Omega+E^{(2)}_\Omega\over V}={p^3_{\rm F-}p^3_{\rm F+}\over9\pi^3}~\!
  {\hbar^2\over m_f}~\!a_0+{2~\!p^3_{\rm F-}p^3_{\rm F+}\over9\pi^4}~\!
  {\hbar^2\over m_f}~\!a^2_0\Lambda+
  {\hbar^2\over m_f}~\!32a_0^2~\! {I\over(2\pi)^7}~\!.\label{eqn:SumE1andE2}
\end{eqnarray}
The regions of integrations over $d^3\mathbf{u}$ and over $d^3\mathbf{t}$ in
$I$ are determined by the intersections of two Fermi spheres of unequal radii,
$p_{\rm F-}$ and $p_{\rm F+}$, the centers of which are displaced from the origin
of the $\mathbf{u}$ (of the $\mathbf{t}$) space by the vectors $-\mathbf{s}$
($\mathbf{s}$ will be taken to determine the $z$-axes of the $\mathbf{u}$ and
$\mathbf{t}$ spaces in the integrals over $d^3\mathbf{u}$ and $d^3\mathbf{t}$)
and $\mathbf{s}$, respectively. The integral over $\mathbf{u}$ is over the
infinite exterior of both spheres and is, therefore, divergent; the
integration over $\mathbf{t}$ covers the intersection of the two spheres. For
this reason the outermost integration over $s\equiv|\mathbf{s}|$ is restricted
to $s\leq s_{\rm max}={1\over2}(p_{\rm F+}+p_{\rm F-})$ because if
$s>{1\over2}(p_{\rm F+}+p_{\rm F-})$, the two spheres
which determine the region of integration over $\mathbf{t}$ become disjoint.
It will be convenient to write  $I=8(2\pi)^3J(p_{\rm F-},p_{\rm F+})$ with
\begin{eqnarray}
  J(p_{\rm F-},p_{\rm F+})=\int_0^{s_{\rm max}}\!ds~\!s^2{1\over4\pi}\!
  \int\!d^3\mathbf{t}~\!\theta(p_{\rm F-}-|\mathbf{t}+\mathbf{s}|)~\!
  \theta(p_{\rm F+}-|\mathbf{t}-\mathbf{s}|)~\!g(t,s)~\!,\label{eqn:Jpp}\\
  g(t,s)\equiv g(|\mathbf{t}|,s)={1\over4\pi}\!\int\!d^3\mathbf{u}~\!
  {\theta(|\mathbf{u}+\mathbf{s}|-p_{\rm F-})~\!
    \theta(|\mathbf{u}-\mathbf{s}|-p_{\rm F+})\over
    \mathbf{t}^2-\mathbf{u}^2+i0}~\!.\phantom{aaaaa}\!\nonumber
\end{eqnarray}

\begin{figure}[]
\begin{center}
\begin{picture}(400,200)(5,0)
\CArc(50,50)(50,0,360)
\Vertex(50,50){2}
\CArc(15,50)(12,0,360)
\Vertex(15,50){2}
\Line(-5,50)(105,50)
\ArrowLine(105,50)(106,50)
\Line(32.5,-5)(32.5,105)
\ArrowLine(32.5,105)(32.5,106)
\Text(-5,100)[]{$a)$}
\CArc(180,50)(50,0,360)
\Vertex(180,50){2}
\CArc(160,50)(30,0,360)
\Vertex(160,50){2}
\Line(125,50)(235,50)
\ArrowLine(235,50)(236,50)
\Line(170,-5)(170,105)
\ArrowLine(170,105)(170,106)
\Text(125,100)[]{$b)$}
\CArc(340,50)(50,0,360)
\Vertex(340,50){2}
\CArc(300,50)(30,0,360)
\Vertex(300,50){2}
\Line(265,50)(395,50)
\ArrowLine(395,50)(396,50)
\Line(320,-5)(320,105)
\ArrowLine(320,105)(320,106)
\Text(265,100)[]{$c)$}
\DashLine(320,50)(280,110){3}
\DashCArc(320,50)(15,0,115){1}
\DashArrowArc(320,50)(15,115,118){1}
\Text(325,58)[]{${\vartheta_0}$}
\end{picture}
\end{center}
\caption{Configurations of the Fermi spheres. Dots mark their centers shifted
  by $\mp s$ from the origin of the space.
  $a)$ $p_{{\rm F}-}/p_{{\rm F}+}<1/3$; in this case,
  if $p_{{\rm F}-}<s<(p_{{\rm F}+}-p_{{\rm F}-})/2$, the smaller sphere is entirely
  in the left half of the space. $b)$ $p_{{\rm F}-}/p_{{\rm F}+}>1/3$; in this
  case part of the smaller
  sphere is always in the right half of the space. $c)$ The spheres intersect
  for $(p_{{\rm F}+}-p_{{\rm F}-})/2<s<(p_{{\rm F}+}+p_{{\rm F}-})$. Marked is the
``critical'' polar angle $\vartheta_0$.}
\label{fig:FermiSpheres}
\end{figure}
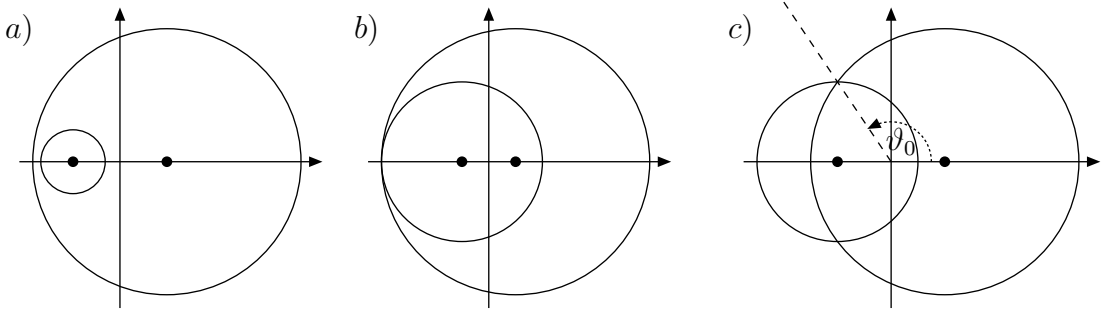

As far as the integral $g(t,s)$ is concerned, the range of the variable $s$
splits into two domains: if $0\leq s\leq s_0={1\over2}(p_{\rm F+}-p_{\rm F-})$,
the sphere of radius $p_{\rm F-}$ is entirely contained inside the one of
radius $p_{\rm F+}$ (see Fig. \ref{fig:FermiSpheres}) and plays no role in
determining the domain of
integration over $\mathbf{u}$: this domain is then just the (infinite)
exterior of the sphere of radius $p_{\rm F+}$ the center of which is at
$u_z=0$, when $s=0$ and moves to the right as $s$ increases but
the origin of the $\mathbf{u}$ space ($u_x=u_y=u_z=0$)
always remains inside this sphere. Thus, in the spherical variables
$u_x=u\sin\vartheta\cos\varphi$, 
$u_x=u\sin\vartheta\sin\varphi$, $u_z=u\cos\vartheta\equiv u~\!\xi$,
the  integration over $du$ is bounded from below by the condition
$u^2-2us~\!\xi-(p^2_{{\rm F}+}-s^2)=0$ and one can write
\begin{eqnarray}
  g(t,s)={1\over2}\int_{-1}^{+1}\!d\xi\!\int_{u(\xi,s)}^\infty\!du~\!{u^2\over
    t^2-u^2+i0}\phantom{aaaaaaaaaaaaaaaaaaaa}\nonumber\\
  =\int^\infty_0\!du~\!{u^2\over t^2-u^2+i0}
  +{1\over2}\int_{-1}^{+1}\!d\xi\!\int^{u(\xi,s)}_0\!du~\!{u^2\over u^2-t^2-i0}~\!,
  \nonumber
\end{eqnarray}
where $u(\xi,s)=s~\!\xi+\sqrt{p^2_{\rm F+}-s^2(1-\xi^2)}$. The first,
divergent, integral is proportional to the integral
(\ref{eqn:I0integralDef}). The second one can be worked out
using the trick
given in Appendix C of \cite{Mloty2}, that is introducing under the
integral over $\xi$ the factor $1=d\xi/d\xi$, taking this integral
by parts and then trading the remaining integration over $\xi$ for
the integration over $u(\xi,s)$. This, upon using the result
(\ref{eqn:I0integralResult}), leads to ($v=u^2(\xi,s)$; terms
of (\ref{eqn:I0integralResult})
vanishing in the limit $\Lambda\rightarrow\infty$ are omitted)
\begin{eqnarray}
  g(t,s)=-i{\pi\over2}~\!t-\Lambda+{1\over2}\left\{
  \int^{u(1,s)}_0\!du~\!{u^2\over u^2-t^2-i0}\right.
  \phantom{aaaaaaaaaaaaaaaaaaaaaaaaaaaa}\nonumber\\
  \left.+\int^{u(-1,s)}_0\!du~\!{u^2\over u^2-t^2-i0}
  -{1\over4s}\!\int_{u^2(-1,s)}^{u^2(1,s)}\!dv~\!
  {v-(p^2_{\rm F+}-s^2)\over v-t^2-i0}\right\}.\nonumber
\end{eqnarray}
The sum of imaginary parts of the three integrals should cancel the
explicit imaginary contribution which resulted from the divergent integral.
In fact, of the three integrals only the first two do develop an imaginary
part.\footnote{Indeed, it is easy to see geometrically that the maximal value
  of $t$ reached in the outer integral is $s+p_{\rm F-}$, whereas the lower limit
  of the third integral is $(p_{\rm F+}-s)^2$; since the function $g(t,s)$ computed
  here is valid only up to $s\leq s_0=(p_{\rm F+}-p_{\rm F-})/2$, the variable $t^2$
  never exceeds the lower limit of integration over $v$.}
We therefore compute these integrals using the standard tricks (the Sochocki
formula)
\begin{eqnarray}
  \int^{u_{\rm max}}_0\!du~\!{u^2\over u^2-t^2-i0}
  =i{\pi\over2}~\!t+u_{\rm max}+{t\over2}~\!P\!\int^{u_{\rm max}}_0\!du\left(
  {1\over u-t}-{1\over u+t}\right)\nonumber\\
  =i{\pi\over2}~\!t+u_{\rm max}+{t\over2}\ln{u_{\rm max}-t\over u_{\rm max}+t}~\!.
  \phantom{aaaaaaaaaaaaaa}\!\label{eqn:intTouMAX}
\end{eqnarray}
Since $u(1,s)=p_{\rm F+}+s$, $u(-1,s)=p_{\rm F+}-s$ one obtains (the imaginary
part of $g(t,s)$ indeed disappears)
\begin{eqnarray}
  g(t,s)=-\Lambda+p_{\rm F+}+{t\over4}\ln{p_{\rm F+}+s-t\over p_{\rm F+}+s+t}
  +{t\over4}\ln{p_{\rm F+}-s-t\over p_{\rm F+}-s+t}
  -{1\over8s}\!\int_{u^2(-1,s)}^{u^2(1,s)}\!dv~\!
  {v-(p^2_{\rm F+}-s^2)\over v-t^2-i0}~\!.\nonumber
\end{eqnarray}
Working out the last integral (ignoring $-i0$, as the pole of the integrand
is outside the range of the integration)
one gets the formula for the function $g(t,s)$ in the range $0\leq s\leq s_0$
\begin{eqnarray}
  g(t,s)=-\Lambda+{1\over2}~\!p_{\rm F+}
  +{t\over4}\ln{(p_{\rm F+}-t)^2-s^2\over(p_{\rm F+}+t)^2-s^2}
  +{p^2_{\rm F+}-s^2-t^2\over8s}\ln{(p_{\rm F+}+s)^2-t^2\over(p_{\rm F+}-s)^2-t^2}~\!.
  \phantom{a}\label{eqn:g2t1}
\end{eqnarray}
Integrated over $t$, $\eta$ and then over $s$ from 0 to $s_0$, it 
gives the corresponding contribution to the ground state energy density.
The integration over $t$ and $\eta$ is in this case ($0<s<s_0$; the smaller
sphere entirely inside the larger one) over the interior of the (smaller)
sphere of radius $p_{\rm F-}$. If $p_{\rm F-}/p_{\rm F+}>1/3$ the origin
$\mathbf{t}=\mathbf{0}$ of the $\mathbf{t}$-space is, as long as $0<s<s_0$,
always inside this sphere (see Fig. \ref{fig:FermiSpheres}b)
and the integrations over $t$ and $\eta$ are given by the expression
\begin{eqnarray}
  \int_{-1}^1\!d\eta\!\int_0^{t(\eta,s)}\!dt~\!t^2~\!g(t,s)~\!,\nonumber
\end{eqnarray}
with $t(\eta,s)=-s~\!\eta+\sqrt{p^2_{\rm F-}-s^2(1-\eta^2)}$.
If, instead $p_{\rm F-}/p_{\rm F+}<1/3$, the  integration domain
$0\leq s\leq s_0$ has to be further split into $0\leq s\leq p_{\rm F-}$ and
$p_{\rm F-}\leq s\leq s_0$. In the first case the integrations over $t$ and
$\eta$ are given by the expression given above. However,
when $p_{\rm F-}\leq s\leq s_0$, the sphere of radius $p_{\rm F-}$ is entirely in
the left half of the $\mathbf{t}$-space (see Fig. \ref{fig:FermiSpheres}a)
and the integrations over $t$ and $\eta$ is given by\footnote{Again it
can be checked that if $g(t,s)=1$, this expression gives $(2/3)p^3_{\rm F-}$.}
\begin{eqnarray}
  \int_{-1}^{-\sqrt{1-p_{\rm F-}^2/s^2}}\!d\eta\!\int_{t_-(\eta,s)}^{t_+(\eta,s)}\!dt~\!
  t^2~\!g(t,s)~\!,\nonumber
\end{eqnarray}
with $t_\mp(\eta,s)=-s\eta\mp\sqrt{p^2_{\rm F-}-s^2(1-\eta^2)}$ (the upper
limit of the integral over $\eta$ is determined by the equality $t_-=t_+$).

Thus if we write the integral (\ref{eqn:Jpp}) as $J=J_1+J_2$ where $J_1$ is
the contribution of the range of $s$ for which the sphere of radius
$p_{\rm F-}$ is entirely inside the one of radius $p_{\rm F+}$, we get
\begin{eqnarray}
  J_1=\int_0^{s_0}\!ds~\!s^2~\!{1\over2}\!\int_{-1}^1\!d\eta\!\int_0^{t_+(\eta,s)}\!dt
  ~\!t^2~\!g(t,s)~\!,\phantom{aaaa}{\rm if}\phantom{aa}p_{\rm F-}>{1\over3}p_{\rm F+}~\!,
  \phantom{aaaa}{\rm and}\phantom{aaaaaaaaaaaaa}~\nonumber\\
  J_1=\int_0^{p_{\rm F-}}\!\!\!\!\!ds~\!s^2~\!{1\over2}
  \!\int_{-1}^1\!d\eta\!\int_0^{t_+(\eta,s)}\!dt
  ~\!t^2~\!g(t,s)+\int_{p_{\rm F}}^{s_0}\!ds~\!s^2~\!{1\over2}\!
  \int_{-1}^{-\sqrt{1-p_{\rm F-}^2/s^2}}\!d\eta\!\int_{t_-(\eta,s)}^{t_+(\eta,s)}\!dt~\!
  t^2~\!g(t,s)~\!,\nonumber
\end{eqnarray}
if $p_{\rm F-}<{1\over3}p_{\rm F+}$. Since when $g(t,s)\equiv1$
both integrations over $\eta$ and $t$
give $(2/3)p_{\rm F-}^3$, the divergent part of $J_1$ is
\begin{eqnarray}
  J^{\rm div}_1={1\over3}~\!s_0^3~\!{1\over3}~\!p_{\rm F-}^3(-\Lambda)
  =-{1\over72}~\!(p_{\rm F+}-p_{\rm F-})^3p_{\rm F-}^3\Lambda~\!.\label{eqn:J1divPart}
\end{eqnarray}

To simplify the evaluation of the (finite part of the) integral $J_1$
one can shift the variable $\mathbf{t}$ by writing
$\mathbf{t}=\mathbf{t}^\prime-\mathbf{s}$ and introducing the
spherical coordinate system in the $\mathbf{t}^\prime$ space with the $t^\prime_z$
axis taken in the direction of the vector $\mathbf{s}$. Then
\begin{eqnarray}
  J_1={1\over2}\int_0^{s_0}\!ds~\!s^2\int_{-1}^1\!d\eta\int_0^{p_{\rm F-}}\!
  dt^\prime~\!t^{\prime2}~\!g\!\left(\sqrt{t^{\prime2}-2t^\prime s\eta+s^2},s\right),
  \nonumber
\end{eqnarray}
without the need to distinguish the cases $p_{\rm F-}>{1\over3}p_{\rm F+}$ and
$p_{\rm F-}<{1\over3}p_{\rm F+}$. Numerical evaluation shows that both ways
of computing $J_1$ yield the same results.\footnote{Yet another, the simplest
  way, of evaluating this contribution (without distinguishing the cases
  $p_{\rm F-}>{1\over3}p_{\rm F+}$ and $p_{\rm F-}<{1\over3}p_{\rm F+}$) is to use
  the Mathematica package instruction\\
  0.5NIntegrate$[s^2t^2g[t,s]~\!{\rm Boole}[t^2+2tsx+s^2<p^2_{\rm F-}],~\!\{s,0,s_0\},
    ~\!\{x,-1,1\},~\!\{t,0,\infty\}]$.}
\vskip0.2cm

We now compute the function $g(t,s)$ for $s_0\leq s\leq s_{\rm max}$ and the
corresponding contribution $J_2$ to the integral (\ref{eqn:Jpp}).
In this regime the two Fermi spheres which determine the ranges of
integrations over $\mathbf{u}$ and over $\mathbf{t}$ intersect one another.
In the $\mathbf{u}$ space the $z$ coordinate $u_z$ of the intersection and
its distance $u_0$ from the origin are determined by  solving the equations
\begin{eqnarray}
  &&u^2_\perp+(u_z-s)^2=p^2_{\rm F+}~\!,\nonumber\\
  &&u^2_\perp+(u_z+s)^2=p^2_{\rm F-}~\!,\phantom{aa}\nonumber
\end{eqnarray}
which give
\begin{eqnarray}
  u^0_z=-{p^2_{\rm F+}-p^2_{\rm F-}\over4s}~\!,\phantom{aaa}u^2_0={1\over2}\left(
  p^2_{\rm F+}+p^2_{\rm F-}\right)-s^2~\!.\nonumber
\end{eqnarray}
In the spherical system the ``critical'' angle $\vartheta_0$ corresponding to the 
intersection of the spheres (marked in Fig. \ref{fig:FermiSpheres}c) is given by
\begin{eqnarray}
  \cos\vartheta_0=\xi_0={u_z^0\over u_0}=-{p^2_{\rm F+}-p^2_{\rm F-}\over4s\sqrt{{1\over2}\left(
  p^2_{\rm F+}+p^2_{\rm F-}\right)-s^2}}~\!.\nonumber
\end{eqnarray}
Therefore, if $s_0\leq s\leq s_{\rm max}$ (i.e. when the two Fermi
spheres intersect), the function $g(t,s)$ is given by
\begin{eqnarray}
  g(t,s)={1\over2}\int_{\xi_0}^1\!d\xi\int_{u_+(\xi,s)}^\infty\!du~\!{u^2\over t^2-u^2+i0}
  +{1\over2}\int_{-1}^{\xi_0}\!d\xi\int_{u_-(\xi,s)}^\infty\!du~\!{u^2\over t^2-u^2+i0}~\!,
  \nonumber
\end{eqnarray}
where now $u_+(\xi,s)=s~\!\xi+\sqrt{p^2_{\rm F+}-s^2(1-\xi^2)}$,
$u_-(\xi,s)=-s~\!\xi+\sqrt{p^2_{\rm F-}-s^2(1-\xi^2)}$; of course
$u_+(\xi_0,s)=u_-(\xi_0,s)\equiv u_0$. (Because the integrals are over
the exterior of the spheres, the fact that the origin of the $\mathbf{u}$
space may not, for some ranges of the parameters, be inside the smaller
sphere of radius $p_{\rm F-}$ is irrelevant here). After extracting the
divergent part of these two integrals as previously,
they combine giving to the integral $-2\pi^2I_0$ and one obtains
\begin{eqnarray}
  g(t,s)=-\Lambda-i{\pi\over2}~\!t+
{1\over2}\int_{\xi_0}^1\!d\xi\int_0^{u_+(\xi,s)}\!du~\!{u^2\over u^2-t^2-i0}
  +{1\over2}\int_{-1}^{\xi_0}\!d\xi\int_0^{u_-(\xi,s)}\!du~\!{u^2\over u^2-t^2-i0}~\!.
  \nonumber
\end{eqnarray}

It is now straightforward to compute $J_2^{\rm div}$ and to check the
cancellation of $\Lambda$. Indeed, the integral
\begin{eqnarray}
  {1\over4\pi}\int\!d^3\mathbf{t}~\!
  \theta(p_{\rm F-}-|\mathbf{t}+\mathbf{s}|)~\!\theta(p_{\rm F+}-|\mathbf{t}-\mathbf{s}|)
  (-\Lambda)~\!,\nonumber
\end{eqnarray}
can be done  using the already computed integrals:
shifting the origin of the coordinate system of the $\mathbf{t}$ space so
that the intersection of the two Fermi spheres occurs at $t_z^\prime=0$
one readily  obtains the result
\begin{eqnarray}
  -{\Lambda\over2}\left\{\left[{1\over3}p^3_{\rm F-}-{1\over2}p^2_{\rm F-}(s-\varepsilon)
    +{1\over6}(s-\varepsilon)^3\right]+
  \left[{1\over3}p^3_{\rm F+}-{1\over2}p^2_{\rm F+}(s+\varepsilon)
    +{1\over6}(s+\varepsilon)^3\right]\right\},\nonumber
\end{eqnarray}
with $\varepsilon=(p^2_{\rm F+}-p^2_{\rm F-})/4s\equiv-u_z^0$.
This should be integrated from $s_0={1\over2}(p_{\rm F+}-p_{\rm F-})$
to $s_{\rm max}={1\over2}(p_{\rm F+}+p_{\rm F-})$ with the weight $s^2$.
Mathematica does it readily and the result is
\begin{eqnarray}
  J_2^{\rm div}=-\Lambda\left({p_{\rm F+}^2p_{\rm F-}^4\over24}-{p_{\rm F+}p_{\rm F-}^5\over24}
  +{p_{\rm F-}^6\over72}\right).\nonumber
\end{eqnarray}
Combining this with the divergent part (\ref{eqn:J1divPart}) of $J_1$ one
gets
\begin{eqnarray}
  J_1^{\rm div}+J_2^{\rm div}=-\Lambda~\!{p_{\rm F+}^3p_{\rm F-}^3\over72}~\!,\nonumber
\end{eqnarray}
which is precisely what is needed to cancel in (\ref{eqn:SumE1andE2}) the
term explicitly proportional to $\Lambda$ which comes from expressing
$C_0$ in terms of the scattering length in the first order result. Thus, as
expected, the divergences disappear when observable quantities (the ground
state energy) computed
within the effective theory are expressed in terms of other observable
quantities (the scattering lengths).
\vskip0.2cm

To work out the imaginary parts of the two remaining integrals we again
use the trick with taking the integral over $\xi$ by parts
after inserting into it $1=d\xi/d\xi$. This gives
\begin{eqnarray}
  \int_{\xi_0}^1\!d\xi\int_0^{u_+(\xi,s)}\!du~\!{u^2\over u^2-t^2-i0}
  =\int_0^{u_+(1,s)}\!du~\!{u^2\over u^2-t^2-i0}
  -\xi_0\!\int_0^{u_+(\xi_0,s)}\!du~\!{u^2\over u^2-t^2-i0}\nonumber\\
  -{1\over4s}\!\int_{u^2_+(\xi_0,s)}^{u^2_+(1,s)}\!dv~\!{v-(p_{\rm F+}^2-s^2)\over
    v-t^2-i0}~\!,\phantom{aaaaaaaaaaaaaaaa}\nonumber
\end{eqnarray}
and, similarly,
\begin{eqnarray}
  \int_{-1}^{\xi_0}\!d\xi\int_0^{u_-(\xi,s)}\!du~\!{u^2\over u^2-t^2-i0}
  =\xi_0\!\int_0^{u_-(\xi_0,s)}\!du~\!{u^2\over u^2-t^2-i0}
  +\int_0^{u_-(-1,s)}\!du~\!{u^2\over u^2-t^2-i0}\nonumber\\
  +{1\over4s}\!\int_{u^2_-(-1,s)}^{u^2_-(\xi_0,s)}\!dv~\!{v-(p_{\rm F-}^2-s^2)\over
    v-t^2-i0}~\!.\phantom{aaaaaaaaaaaaaaaa}\nonumber
\end{eqnarray}
Since $u_+(\xi_0,s)=u_-(\xi_0,s)=u_0$,  the
terms explicitly proportional to $\xi_0$ mutually cancel out,
when the two integrals are added, and the sum reads
\begin{eqnarray}
  \int_0^{p_{\rm F+}+s}\!du~\!{u^2\over u^2-t^2-i0}
  +\int_0^{p_{\rm F-}+s}\!du~\!{u^2\over u^2-t^2-i0}
  \phantom{aaaaaaaaaaaaaaaaaaaaaaaaa}\nonumber\\
  -{1\over4s}\!\int_{u^2_0}^{(p_{\rm F+}+s)^2}\!dv~\!{v-(p_{\rm F+}^2-s^2)\over
    v-t^2-i0}
  +{1\over4s}\!\int_{(p_{\rm F-}+s)^2}^{u^2_0}\!dv~\!{v-(p_{\rm F-}^2-s^2)\over
    v-t^2-i0}~\!.\nonumber
\end{eqnarray}
The first two integrals do have imaginary parts which can be extracted
using the Sochocki formula that is, using the result (\ref{eqn:intTouMAX});
their sum (the above expression enters $g(t,s)$ divided by two) precisely
cancels the imaginary part which arose from the divergent integral.
The result then is (it is clear that $u_0<p_{{\rm F}\pm}+s$, so
the limits of the last integral are  interchanged)
\begin{eqnarray}
  g(t,s)=-\Lambda+{1\over2}(p_{\rm F+}+p_{\rm F-}+2s)
  +{t\over4}\ln{p_{\rm F+}+s-t\over p_{\rm F+}+s+t}
  +{t\over4}\ln{p_{\rm F-}+s-t\over p_{\rm F-}+s+t}\phantom{aaaaaa}\nonumber\\
  -{1\over8s}\!\int_{u^2_0}^{(p_{\rm F+}+s)^2}\!dv~\!{v-(p_{\rm F+}^2-s^2)\over
    v-t^2-i0}
  -{1\over8s}\!\int_{u^2_0}^{(p_{\rm F-}+s)^2}\!dv~\!{v-(p_{\rm F-}^2-s^2)\over
    v-t^2-i0}~\!.\nonumber
\end{eqnarray}
The final step is to work out the remaining integrals. They are similar
to the one already computed and one gets
\begin{eqnarray}
  g(t,s)=-\Lambda+{1\over4}(p_{\rm F+}+p_{\rm F-}+2s)
  +{t\over4}\ln{p_{\rm F+}+s-t\over p_{\rm F+}+s+t}
  +{t\over4}\ln{p_{\rm F-}+s-t\over p_{\rm F-}+s+t}\phantom{aaaaaaaa}\nonumber\\
 +~\!{p^2_{\rm F+}-t^2-s^2\over8s}\ln{(p_{\rm F+}+s)^2-t^2\over u_0^2-t^2}
 +{p^2_{\rm F-}-t^2-s^2\over8s}\ln{(p_{\rm F-}+s)^2-t^2\over u_0^2-t^2}~\!.\phantom{a}
 \label{eqn:g2ts}
\end{eqnarray}
In the limit $p_{\rm F-}=p_{\rm F+}=k_{\rm F}$
this goes over into the function
\begin{eqnarray}
  g(t,s)=-\Lambda+{1\over2}~\!(k_{\rm F}+s)+{t\over2}\ln{k_{\rm F}+s-t\over k_{\rm F}+s+t}
  +{k^2_{\rm F}-s^2-t^2\over4s}\ln{(k_{\rm F}+s)^2-t^2\over k^2_{\rm F}-s^2-t^2}~\!.\nonumber
\end{eqnarray}
arising in the case of equal densities of spin up and spin down fermions.
In this case, in which the integral $J_1$ is zero, the remaining integrals over $s$,
$\eta$ and $t$ which give $J_2=J$ can be even worked out explicitly \cite{HamFur00}
with the result
\begin{eqnarray}
  J(k_{\rm F},k_{\rm F})=-{1\over72}~\!\Lambda k^6_{\rm F}
  +k^7_{\rm F}~\!{11-2\ln2\over24\cdot35}~\!.\nonumber
\end{eqnarray}

\begin{figure}
\psfig{figure=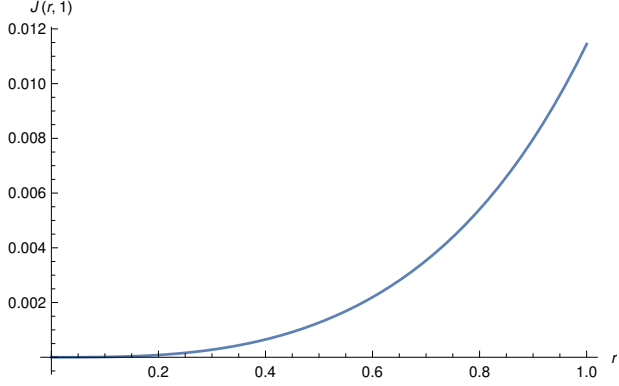,width=8.cm,height=5.0cm} 
\caption{Plot of the function $J(r,1)$. The value
  $J(1,1)=0.0114449=(11-2\ln2)/840$ is the the result
  of \cite{HamFur00}.}
\label{fig:Jplot}
\end{figure}

\begin{figure}
\psfig{figure=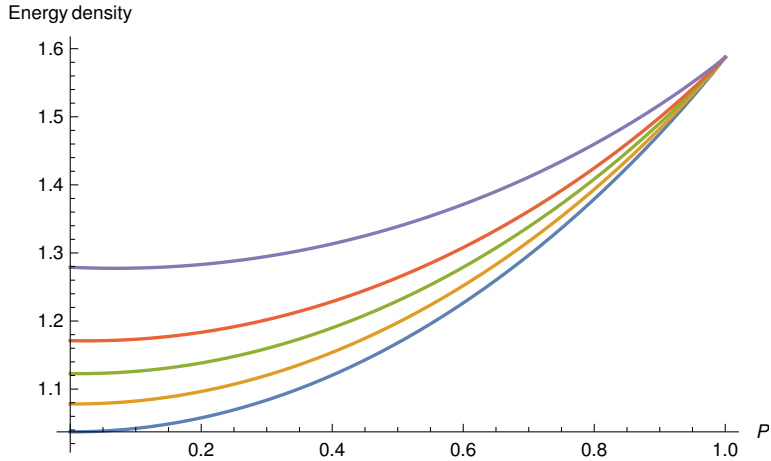,width=10.cm,height=6.0cm} 
\caption{Energy density $E_\Omega/V$ in units $(3/5)\hbar^2k^5_{\rm F}/6\pi^2m_f
  =(N/V)(\hbar^2k^2_{\rm F}/2m_f)(3/5)$
  of the polarized gas of spin $1/2$ fermions as a function of
  its polarization $P=(N_+-N_-)/N$  for different values
  (from below) 0.1 (blue), 0.2 (yellow), 0.3 (green), 0.4 (red)
  and 0.6 (blue) of the expansion parameter $k_{\rm F}a_0$.}
\label{fig:Energy}
\end{figure}

If $p_{\rm F-}<p_{\rm F+}$, the integrals $s$, $\eta$ and $t$ can be easily
evaluated numerically.\footnote{The simplest way
  is to use the Mathematica numerical integration routine
  to integrate over the domain $s_0\leq s\leq s_{\rm max}$, $-1\leq x\leq1$,
  $0\leq t\leq\infty$, imposing the conditions $t^2+2tsx+s^2<p^2_{\rm F-}$
  and $t^2-2tsx+s^2<p^2_{\rm F+}$ but we have also evaluated it using
  other methods always with the same results.}  
It is convenient to write the complete function
$J(p_{\rm F-},p_{\rm F+})$ defined in (\ref{eqn:Jpp}), setting $\Lambda=0$
as $p_{\rm F+}^7J(r,1)$ with $0\leq r\equiv p_{\rm F-}/p_{\rm F+}\leq 1$.
The function $J(r,1)$ is shown in Fig. \ref{fig:Jplot}.
The complete result can be therefore written as
\begin{eqnarray}
  {E_\Omega\over V}={\hbar^2p_{\rm F+}^2\over m_f}~\!{p_{\rm F+}^3\over6\pi^2}
  \left\{{3\over10}\left(1+r^5\right)
  +{2\over3\pi}~\!r^3\left(p_{\rm F+}a_0\right)
  +{96\over\pi^2}\left(p_{\rm F+}a_0\right)^2J(r,1)+\dots\right\}.
\end{eqnarray}
It is, however better to express it in terms of $k_{\rm F}=(3\pi^2\rho)^{1/3}$,
where $\rho=N/V$ - the Fermi wave vector in the case $N_\uparrow=N_\downarrow=N/2$,
which does not change when the ratio $r=p_{\rm F-}/p_{\rm F+}$ (i.e. the system's
polarization) is varied. Since $p_{\rm F+}=k_{\rm F}(2/(1+r^3))^{1/3}$,
\begin{eqnarray}
  {E_\Omega\over V}={\hbar^2k_{\rm F}^2\over m_f}~\!{k_{\rm F}^3\over6\pi^2}
  \left({2\over1+r^3}\right)^{5/3}\left\{{3\over10}\left(1+r^5\right)
  +{2\over3\pi}~\!r^3\left({2\over1+r^3}\right)^{1/3}\left(k_{\rm F}a_0\right)
  \right.\phantom{aaaaaaa}\nonumber\\
  \left.+{96\over\pi^2}\left({2\over1+r^3}\right)^{2/3}
  \left(k_{\rm F}a_0\right)^2J(r,1)+\dots\right\}.\label{eqn:Final}
\end{eqnarray}
This energy density is plotted in Fig. \ref{fig:Energy} as a function of the
system's polarization $P=(N_+-N_-)/N$ ($0\leq P\leq1$; it is related to the
variable $r$ by $r=[(1-P)/(1+P)]^{1/3}$) for five different values of the
expansion parameter $k_{\rm F}a_0$. We have checked that our result
agrees with that of \cite{KANNO}.\footnote{The precise relation, checked
  numerically, of the
  function $I(P)$ used in \cite{KANNO} (the variable $r$ used there
  is our polarization $P$) to the function $J$ defined by (\ref{eqn:Jpp})
  evaluated with the cutoff $\Lambda=0$ and $k_{\rm F}=1$ is
  \begin{eqnarray}
    I(P)=160~\!(1+P)^{7/3}J(r(P),1)~\!,\phantom{aaa}r(P)=((1-P)/(1+P))^{1/3}.
    \nonumber
  \end{eqnarray}
  }
The computed correction of order
$(k_{\rm F}a_0)^2$ is rather small, $\sim1$\%  for $k_{\rm F}a_0=0.2$ and $r=1$
and decreases with decreasing $r$. All curves assume at $P=1$ the same value
$2^{2/3}=1.5874$ - due to the Pauli exclusion principle interactions do not
induce any corrections to the ground state of a fully polarized ($P=1$, $r=0$)
system of fermions.\footnote{This readily follows from the form of the effective
  iteraction written in terms of the field operators $\phi_\pm$, $\psi^\dagger_\pm$
  introduced in (\ref{eqn:psiplusminus}) and the absence of the ``sea''
  of oppositely polarized fermions; the same can be also inferred by
  writing explicitly the original interation Hamiltonian (\ref{eqn:ModelInteraction})
  in terms of creation and annihilation operators associated with
  different spin projections.}
Note also that the prefactor
$\hbar^2k^5_{\rm F}/6\pi^2m_f$ in (\ref{eqn:Final}) can  be written in the
form $(N/V)(\hbar^2k^2_{\rm F}/2m_f)$. Therefore, our Fig. \ref{fig:Energy} can
be directly compared to Fig. 3 of ref. \cite{QMC10}: our curve for
$k_{\rm F}a_0=0.6$ corresponds to the lowest curve in this plot obtained
for a model repulsive potential by a numerical estimate of the exact ground
state energy. It is seen that the result of the second order expansion
is somewhat lower than the numerical estimate. This agrees with
the comparison of the ground state energies of the unpolarized system
($P=0$, $r=1$) performed in Fig. 2 of ref. \cite{QMC10} which shows that 
the perturbative expansion of the ground state energy in powers of
$k_{\rm F}a_0$ remains reliable up to $k_{\rm F}a_0\simlt0.5$ but is
systematically below the numerical estimates of the exact value.

\section{Summary}
\label{sec:summary}

We have recomputed the order $(k_{\rm F}a_0)^2$, where $a_0$ is the $s$-wave
scattering length and $k_{\rm F}=(3\pi^2N/V)^{1/3}$, correction to the ground
state energy of a polarized gas of (nonrelativistic) fermions of spin $1/2$
using the effective theory approach proposed in \cite{HamFur00} which does
not require specifying explicitly the (spin independent) interaction potential. 
We have demonstrated the cancellation of ultraviolet divergences
when the result is expressed in terms of the scattering length.
Our result obtained by the method applicable to arbitrary repulsive
interaction potentials is identical with that of \cite{KANNO} obtained with
the help of traditional methods within the specific model of hard spheres.
That it should be so is almost obvious in the effective theory approach
but wasn't such in the old framework.

Since the main technical problem of this approach is only isolating ultraviolet
divergences and working out cancellation of imaginary contributions,
it seems that with some more labour the computations presented here
could be extended to yet higher orders of the expansion, similarly as
was done in the case of unpolarized system in \cite{HamFur00} and
\cite{WeDrSch}. A more challenging task would be obtaining a rigorous
estimate of the high order terms of the perturbative expansion which
could allow to assess the range of its convergence.

In this paper we have considered the polarized diluted gas of
(nonrelativistic) interacting spin $1/2$ fermions, working 
in the continuum version of the theory. Our results can be most
naturally applied to atomic gases bound in traps. An analogous problem
can of course be also formulated using the lattice version,
that is within the paradigmatic Hubbard model, with obvious
applications to atomic gases bound in periodic laser traps and to
the solid state systems. As far as we know, there are no second
order results similar to ours in this other version (rigorous first order
results have been given in \cite{Giuliani} and \cite{SeiringerYin})
and it would be interesting to try to obtain them.

\vskip0.5cm
\noindent{\bf Acknowledgments.} We would like to thank Pierbiagio Pieri
for bringing reference \cite{KANNO} to our attention.

\end{document}